\documentclass[twocolumn]{aastex63}

\usepackage{amsmath}
\usepackage{enumitem}
\usepackage{multirow}
\usepackage{hyperref}
\usepackage{lineno}
\renewcommand{\edit}[1]{{{#1}}}

\defcitealias{Lam:2020}{L20}
\defcitealias{Mroz:2019}{M19}

\received{}
\revised{}
\accepted{}
\submitjournal{}

\shorttitle{Photometric Microlensing and the IFMR} 
\shortauthors{Rose, Lam, Lu}

\begin{document}

\title{The Impact of Initial-Final Mass Relations on Black Hole Microlensing}

\author[0000-0003-4725-4481]{Sam Rose} 
\correspondingauthor{Sam Rose}
\email{srose@caltech.edu}
\affiliation{University of California, Berkeley, Department of Astronomy, Berkeley, CA 94720}
\affiliation{\textcolor{black}{California Institute of Technology, Cahill Center for Astronomy and Astrophysics, Pasadena, CA 91125}}
\author[0000-0002-6406-1924]{Casey Y. Lam}
\correspondingauthor{Casey Y. Lam}
\email{casey\_lam@berkeley.edu}
\affiliation{University of California, Berkeley, Department of Astronomy, Berkeley, CA 94720}
\author[0000-0001-9611-0009]{Jessica R. Lu}
\affiliation{University of California, Berkeley, Department of Astronomy, Berkeley, CA 94720}
\author[0000-0002-7226-0659]{Michael Medford}
\affiliation{University of California, Berkeley, Department of Astronomy, Berkeley, CA 94720}
\author[0000-0003-2874-1196]{Matthew W. Hosek Jr.}
\altaffiliation{Brinson Prize Fellow}
\affiliation{University of California, Los Angeles, Department of Astronomy, Los Angeles, CA 90095}
\author[0000-0002-0287-3783]{Natasha S. Abrams}
\affiliation{University of California, Berkeley, Department of Astronomy, Berkeley, CA 94720}
\author[0000-0003-0447-8426]{Emily Ramey}
\affiliation{University of California, Berkeley, Department of Astronomy, Berkeley, CA 94720}
\author[0000-0002-4951-8762]{Sergiy S. Vasylyev}
\affiliation{University of California, Berkeley, Department of Astronomy, Berkeley, CA 94720}

\begin{abstract}
Uncertainty in the initial-final mass relation (IFMR) has long been a problem in understanding the final stages of massive star evolution. 
One of the major challenges of constraining the IFMR is the difficulty of measuring the mass of non-luminous remnant objects (i.e. neutron stars and black holes). 
Gravitational wave detectors have opened the possibility of finding large numbers of compact objects in other galaxies, but all in merging binary systems. 
Gravitational lensing experiments using astrometry and photometry
\edit{are}
capable of finding compact objects, both isolated and in binaries, in the Milky Way.
In this work we improve the \texttt{PopSyCLE} microlensing simulation code in order to explore the possibility of constraining the IFMR using the Milky Way microlensing population. 
\edit{We predict that the Roman Space Telescope's microlensing survey will \textcolor{black}{likely} be able to distinguish different IFMRs based on the differences at the long end of the Einstein crossing time distribution and the small end of the microlensing parallax distribution\textcolor{black}{, assuming the small ($\pi_E \lesssim 0.02$) microlensing parallaxes characteristic of black hole lenses are able to be measured accurately}.
\textcolor{black}{We emphasize that future microlensing surveys need to be capable of  characterizing events with small microlensing parallaxes in order to place the most meaningful constraints on the IFMR.}}
\end{abstract}

\section{Introduction}
\label{sec:Introduction}
The initial-final mass relation (IFMR) maps the initial mass of stars on the main sequence to the mass of their compact remnants.
The form of the IFMR is not well determined due to limitations in both theory and observation, and thus represents a very active area of research in stellar physics (\citet{Lu:2019}, \citet{Costa:2021}, \citet{Heger:2003}, \citet{Ertl:2016}). 
The final mass of a compact object depends not only on its zero age main sequence (ZAMS) mass, but also on factors such as rotation (which causes mixing in a star), metallicity (which governs mass loss rates due to stellar winds, particularly during the post main sequence), multiplicity (stars in close binaries can evolve differently), and core structure just prior to explosion (which can determine what type of compact object will be formed; \citet{Sukhbold:2018} and references therein).

While the IFMR for low mass stars is
\edit{measured using} direct observation of white dwarfs \citep{Cummings:2018, Kalirai:2008} the IFMR for high mass stars \edit{is an open question, due to difficulty in obtaining mass measurements of massive compact remnants.}
While it is possible to measure the mass of some isolated, young neutron stars using pulsar timing \citep{Lorimer:2008} or measure the mass of neutron stars in X-ray binaries \citep{Steiner:2013}, obtaining mass measurements of a large number of isolated stellar mass black holes presents an even greater challenge. 

One method to increase the number of mass measurements for dark, isolated compact objects is to use gravitational microlensing. 
When \edit{a massive} object passes in front of a luminous background source, a transient brightening \edit{and positional shift of the background star} occurs.
While photometric observations are sufficient to detect microlensing events, precise astrometric observations of the centroid shift during the microlensing event are required to break degeneracies and measure the lens mass directly \citep{Lu:2016, Sahu:2017}. 
\textcolor{black}{The first detection of an isolated dark object using photometric and astrometric microlensing was recently reported; the analysis of \citet{Lam:2022} suggests the object is a neutron star or low-mass black hole, while the analysis of \citet{Sahu:2022, Mroz:2022} suggest a black hole is the only possibility}.

\edit{To date, thousands of photometric microlensing events have been detected by dedicated microlensing surveys such as the Optical Gravitational Lensing Experiment \citep[OGLE,][]{Udalski:1992} and the Microlensing Observations in Astrophysics \citep[MOA,][]{Muraki:1999}, as well as other astronomical surveys such as the Zwicky Transient Facility \citep[ZTF,][]{Bellm:2019} \textcolor{black}{and Gaia \citep{Wyrzykowski:2022}}.
However, only a few tens of these photometric events have also been observed astrometrically, as the follow up process is extremely resource intensive and few facilities have the requisite precision.
In the absence of mass measurements due to lack of astrometric microlensing measurements, statistical constraints placed on the IFMR from photometric microlensing alone are the best way to compare our theoretical models for late-stage stellar evolution and stellar death with observation.}

This work explores whether or not photometric microlensing surveys will be able to place meaningful constraints on the IFMR.
In \S \ref{The PopSyCLE Simulation} we discuss the Population Synthesis for Compact object Lensing Events (\texttt{PopSyCLE}) microlensing simulation and the modifications that we have made to it for this work. 
In \S \ref{sec:The SPISEA IFMR Object} we discuss the implementation and characteristics of different IFMRs added to the Stellar Population Interface for Stellar Evolution and Atmospheres (\texttt{SPISEA}) simple stellar population synthesis code.
In \S \ref{sec:Results} we present our findings on the effect of the different IFMRs on the black hole microlensing population (\S \ref{sec:BH Microlensing Statistics}), and whether or not these differences are detectable with OGLE (\S \ref{sec:Constraining the IFMR with OGLE}) or the upcoming Roman Space Telescope's microlensing survey (\S \ref{sec:Constraining the IFMR with the Roman Space Telescope}). 
\edit{In \S \ref{sec:Discussion} we compare the Galactic BH distribution predicted by the different IFMRs to the extragalactic BH distribution detected via gravitational wave mergers, as well as discuss further enhancements to \texttt{PopSyCLE}.
We finish in \S \ref{sec:Conclusions} with a summary of our main conclusions.} 

\section{The \texttt{PopSyCLE} Simulation}
\label{The PopSyCLE Simulation}
\texttt{PopSyCLE} is a microlensing population synthesis tool for the Milky Way \citep{Lam:2020}.
Given a survey location and area as well as other parameters like the length of the survey, number of observations, reddening law, and filter, \texttt{PopSyCLE} will return a list of observable microlensing events and the parameters associated with them (e.g. Einstein crossing time, microlensing parallax, magnitude of brightening).
In this work we make several modifications to \texttt{PopSyCLE} in order to explore the effect of different IFMRs on the Milky Way black hole microlensing population which are described in detail below. 
For a full description of \texttt{PopSyCLE}, see \citet{Lam:2020}. 

\subsection{Milky Way Models using \texttt{Galaxia}}
\label{sec:Milky Way Models using Galaxia}
\texttt{Galaxia} is a resolved stellar simulation of the Milky Way \citep{Sharma:2011}, which serves as the foundation for stellar population synthesis in \texttt{PopSyCLE}.
Given a survey area and location \texttt{Galaxia} will return all the stars located in the conical volume of the projected circular survey area centered on the specified coordinates.
Compact objects are \emph{not} included in the output of \texttt{Galaxia}. 
Each star returned by \texttt{Galaxia} has a position, velocity, mass, age, metallicity, among other parameters.

The \texttt{Galaxia} stellar simulation is based on the Besan\c{c}on analytic model for the Milky Way \citep{Robin:2003}, with a modified version of the disk kinematics that adjusts the velocity in the azimuthal direction \citep{Shu:1969}. 
For a summary of the relevant distributions from which the stellar parameters are drawn for various populations of stars and a brief description of each population, see Tables 1-3 in \citet{Sharma:2011}. 

In the \texttt{PopSyCLE} simulations presented in this paper, we use \edit{the ``v3" Galactic model described in Appendix A of \citet{Lam:2020}, which differs from the default Galactic model of \texttt{Galaxia}.}
\edit{The most salient change is to the Galactic bar.}
The angle of the line connecting the Sun and Galactic Center $\alpha$ and the major axis scale length of the Galactic bar $x_{0}$ are changed from $\alpha$ = 11.1$^{\circ}$ to $\alpha$ = 28$^{\circ}$ and $x_{0}$ = 1.59 kpc to $x_{0}$ = 0.7 kpc \citep{Wegg:2013, Wegg:2015}.
All simulations discussed in \edit{the main text of} this paper are run using these parameters.

\subsection{Compact Object Synthesis using \texttt{SPISEA}}
\label{sec:Compact Object Syntheis using SPISEA}
\texttt{Galaxia} is a stellar survey, which means it does \emph{not} include compact objects. In order to include compact objects in our microlensing simulation we must inject them from another source.
\texttt{SPISEA}\footnote{\texttt{SPISEA} was formerly called \texttt{PyPopStar}, and is referred to as such in \citet{Lam:2020}.} 
\citep{Hosek:2020} is a software package which generates a single age, single metallicity stellar population (i.e. star cluster) based on user controlled parameters such as the initial mass function (IMF), evolution models, atmosphere models, extinction maps, and multiplicity distributions. 
The MESA Isochrones and Stellar Tracks (MIST) stellar evolution models \citep{Choi:2016} are used to evolve the \texttt{SPISEA} clusters to determine which progenitors have left the post-main sequence and become compact objects.
\edit{The IFMR is then used to assign a remnant type (WD, NS, BH) and mass to the compact objects.}
For details on the \texttt{SPISEA} input parameters used in the \texttt{PopSyCLE} simulation see \S 2.2 of \citet{Lam:2020}. 

In the \texttt{SPISEA} code the user can choose the IFMR used. 
Prior to this work, \texttt{SPISEA} contained a single IFMR object, hereafter referred to as Raithel18, based on \citet{Raithel:2018} for black holes and neutron stars and \citet{Kalirai:2008} for white dwarfs.

This work adds two additional IFMR objects to \texttt{SPISEA}\footnote{\edit{Available in \texttt{SPISEA} v2.1 and later.}}. 
One, called Spera15, is based on the Stellar EVolution $N$-body (\texttt{SEVN}) code\footnote{http://web.pd.astro.it/mapelli/group.html} 
\citep{Spera:2015} which is described in \S \ref{sec:The Spera15 IFMR Function}. 
The Spera15 IFMR is a function of progenitor metallicity as well as ZAMS mass, but does not take into account models of explosion physics like \citet{Raithel:2018}. 
The other new IFMR, based on simulations by \citet{Sukhbold:2016} and \citet{Sukhbold:2014}, is called SukhboldN20 and is described in \S \ref{sec:Sukhbold N20 IFMR}. 
The SukhboldN20 IFMR includes metallicity dependence and the explosion physics of \citet{Raithel:2018}.
Both new IFMR objects also use the \citet{Kalirai:2008} white dwarf (WD) IFMR described in \S \ref{sec:The WD IFMR} for low ZAMS mass stars.

\subsection{Metallicity Binning}
\label{sec:Metallicity Binning}
In the original \texttt{PopSyCLE} simulation, to perform the population synthesis of compact objects using \texttt{SPISEA}, the stars from \texttt{Galaxia} are binned according to population (i.e. thin disk, thick disk, bulge) and age (see \S 3 of \citet{Lam:2020} for more details).
For this work, each age bin is further divided by metallicity. 
This is necessary as \texttt{SPISEA} can create only single-age, single-metallicity populations. 
For \texttt{PopSyCLE} runs using the Raithel18 IFMR, which has no metallicity dependence, the only metallicity sub-bin is at solar metallicity and encompasses the full range of metallicities produced by \texttt{Galaxia}. 

\begin{figure}
    \centering
    \includegraphics[scale=0.4]{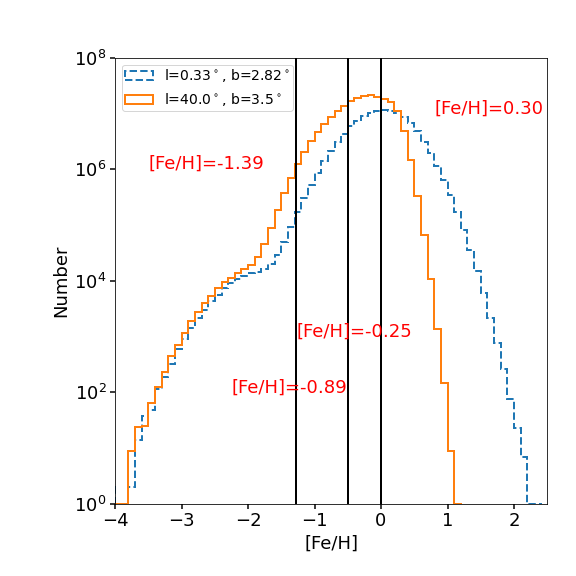}
    \caption{The distribution of [Fe/H] at two different \texttt{Galaxia} field locations.
    Vertical black lines demarcate the chosen metallicity binning for the \texttt{SPISEA} clusters run with a metallicity dependent IFMR, with the metallicity assigned to each bin labeled in red. 
    Note that for the Raithel18 IFMR all \texttt{SPISEA} clusters are solar metallicity only.}
    \label{feh_bins_fig}
\end{figure}

In order to determine the most appropriate bins for simulations using the Spera15 and SukhboldN20 IFMRs,  we first look at the metallicity distribution at two different pointings in the Milky Way using \texttt{Galaxia} (Figure \ref{feh_bins_fig}). 
\edit{We chose metallicity bins such that the number of distinct metallicity isochrones required by \texttt{SPISEA} (which must be generated and stored in order to perform population synthesis) is minimized in order to save on disk space and computational time, and such that the distribution of BH masses given by the Spera15 IFMR did not have arbitrary mass gaps introduced by the metallicity binning.} 

Based on these constraints, we have chosen four metallicity bins for use with the Spera15 and SukhboldN20 IFMRs. 
The first bin contains stars with [Fe/H]~$ < -1.279$ and \edit{are assigned to have values} [Fe/H]~$ = -1.39$ \edit{in the population synthesis}. 
The second bin is -1.279 $<$ [Fe/H] $<$ -0.500 and \edit{stars are assigned to have} [Fe/H] = -0.89. 
The third bin is -0.500 $<$ [Fe/H] $<$ 0.00 and \edit{stars are assigned to have} [Fe/H] = -0.25. 
The final bin includes all stars with [Fe/H] $> 0.00$ (solar metallicity) and \edit{stars are assigned to have} [Fe/H] = 0.30. 
\edit{In future work, modifications made to the \texttt{SPISEA} population synthesis code will allow for interpolation between isochrones, which \texttt{PopSyCLE} will be able to take advantage of with finer metallicity binning.}

\subsection{NS/BH Birth Kick Velocities}
\label{sec:NS/BH Birth Kick Velocities}
One other change that has been made to the original \texttt{PopSyCLE} simulation is the addition of a more \edit{realistic} distribution of birth kick velocities for neutron stars and black holes.
Birth kick velocities are an additional velocity resulting from asymmetries in supernovae explosions, where excess mass loss in one direction \edit{or anisotropic neutron emission} will result in an additional velocity in a random direction for the compact object left behind by the explosion \citep{Janka:1994, Kusenko:1996, Tamborra:2014}. 

In the original version of \texttt{PopSyCLE}, each black hole or neutron star was assigned a constant birth kick velocity (350 km/s for neutron stars and 100 km/s for black holes) in a random direction, which was then added to the existing ``stellar" velocity assigned based on the distribution of stellar velocities coming from \texttt{Galaxia} (see \S 5 of \citet{Lam:2020}). 
The value chosen for the neutron star kick velocity was the average of a Maxwellian distribution reported based on observations of pulsar proper motions in \citet{Hobbs:2005}. 
\edit{In the current version of \texttt{PopSyCLE}, we now implement a more realistic Maxwellian kick distribution for the NS and BH populations, instead of applying a single-valued kick as before.}

Because neutron star birth kicks follow a Maxwellian distribution, we \edit{might} also expect black hole birth kick velocities to be Maxwellian. 
The Maxwellian birth kick velocity distribution used in the simulations for this paper have averages which match the original \texttt{PopSyCLE} values for neutron star and black hole birth kick velocities respectively (i.e. the average of the randomly assigned values for neutron star kick velocity drawn from the Maxwellian distribution was chosen to be 350 km/s, while the average of the kick velocities assigned to the black holes was chosen to be 100 km/s). 

\section{The \texttt{SPISEA} IFMR Object}
\label{sec:The SPISEA IFMR Object}

\subsection{The WD IFMR}
\label{sec:The WD IFMR}
The WD IFMR used in this work is based on \citet{Kalirai:2008}. 
This IFMR is used in all of the \texttt{SPISEA} IFMR objects to derive WD masses for low ZAMS mass progenitors, except those young, luminous WDs already included in the MIST models. 
The \citet{Kalirai:2008} WD IFMR is empirically determined based on observational data in the initial mass range $1.16 M_\odot < M_{ZAMS} < 6.5 M_\odot$, and is given by:
\begin{equation}
\label{eq:wd_ifmr}
    M_{WD} = (0.109\; M_{ZAMS} + 0.394)\; M_\odot.
\end{equation}

We extend the range of this IFMR to $0.5 M_\odot < M_{ZAMS} < 9 M_\odot$ for the Raithel18 and SukhboldN20 IFMR objects and to $0.5 M_\odot < M_{ZAMS} < 7 M_\odot$ for the Spera15 IFMR object, where the upper mass range is chosen such that the IFMR is defined for all ZAMS masses below the lower limit of the NS/BH IFMRs. 
The lower mass range is chosen to match the lower mass limit of the MIST model objects. 
See \S 2.2.2 of \citet{Lam:2020} for more details. 

\subsection{The Raithel18 IFMR}
\label{sec:The Raithel18 IFMR Function}
The original \texttt{PopSyCLE} simulation as described in \citet{Lam:2020} implemented a BH/NS IFMR based on \citet{Raithel:2018} as well as the WD IFMR from \citet{Kalirai:2008} described above in \S \ref{sec:The WD IFMR}. 
The model includes a stochastic process to determine whether or not a black hole or neutron star is ultimately formed from each progenitor, but does not include metallicity dependence. 
Because the simulations used to produce the Raithel18 IFMR assumed solar metallicity, all compact objects are assumed to be have solar metallicity progenitor stars in the original \texttt{PopSyCLE} simulation. 
For a full set of the equations and a description of how the Raithel18 IFMR was implemented see Appendix C of \citet{Lam:2020}.

As the Raithel18 IFMR is based entirely on progenitors stars with solar metallicity, it is missing the most massive compact objects formed from the low metallicity population of stars in the Milky Way \citep{Meng:2008}. 
The masses of BHs formed using the Raithel18 IFMR ranges from around $5 M_{\odot}$ to $16 M_{\odot}$.

In the original \texttt{PopSyCLE} simulation every neutron star produced by the Raithel18 IFMR was assumed to have a mass of 1.6 $M_{\odot}$.
\edit{In the version of \texttt{PopSyCLE} used for this paper, the mass distribution of neutron stars produced by the Raithel18 IFMR is instead drawn from a Gaussian distribution with average 1.36 $M_{\odot}$ and standard deviation 0.09 $M_{\odot}$, based on a compilation of neutron star masses from several observational studies (Appendix \ref{sec:Neutron Star Mass Distribution}).}

\subsection{The Spera15 IFMR}
\label{sec:The Spera15 IFMR Function}
The IFMR is dependent on the metallicity of progenitor stars as well as on their ZAMS mass \citep{Heger:2003, Meng:2008}. 
The Stellar EVolution N-body (\texttt{SEVN}) code is a a software package that models late stage stellar evolution and supernovae physics including the effects of metallicity dependent mass loss \citep{Spera:2015}.
This software, in addition to the stellar evolution models of the PAdova and TRieste Stellar Evolution Code \citep[\texttt{PARSEC},][]{Bressan:2012, Bressan:2013, Tang:2014, Chen:2014, Chen:2015} were used by \citet{Spera:2015} to create an analytical formula for stellar remnant mass as a function of ZAMS mass and the mass fraction of metals $Z$. 
This analytical formula can be found in Appendix C of \citet{Spera:2015}. 

Note that the Spera15 IFMR takes as its argument the mass fraction in metals $Z$ rather than [Fe/H] as returned by \texttt{Galaxia}. 
To convert between Z and [Fe/H] we use the equation
\begin{equation}
    Z = Z_{\odot}10^{[Fe/H]}
\end{equation}
where Z$_{\odot}$ = 0.014 \citep{Ekstrom:2012}. 

We used this analytical formula to create an IFMR object in \texttt{SPISEA}, hereafter referred to as Spera15. 
The function is defined for $M_{ZAMS} \geq 7 M_{\odot}$ and all metallicities. 
It works by first calculating the core mass based on progenitor ZAMS mass and metallicity, and then uses the core mass to calculate the final remnant mass. 
Only once the remnant mass has been calculated is the object determined to be a BH, NS, or WD. 
We use the Chandrasekhar mass of 1.4 $M_{\odot}$ as the limit for WD masses, i.e.~all Spera15 remnants with mass less than 1.4 $M_{\odot}$ are assigned to be WDs. 
All Spera15 remnants with masses between 1.4 $M_{\odot}$ and 3 $M_{\odot}$ are assigned to be NSs, where the upper limit of NS mass is roughly estimated from \citet{Ozel:2016}. 
All Spera15 remnants with mass greater than 3 $M_{\odot}$ are assigned to be BHs. 
\edit{One issue with this method of assigning remnant types is that the outcome is continuous and deterministic based on initial mass.}
Several simulations, including those by \citet{Sukhbold:2016}
suggest \edit{that there is no mass above which stars become BHs and below which they become NSs.
In addition, this method does not produce neutron stars with masses less than the Chandrasekhar mass, when in reality a significant fraction of neutron stars are less than $1.4 M_\odot$.
The result is that the average mass of the compact objects generated is too high.}

Because the Spera15 IFMR includes low metallicity remnant populations, it will return more massive BHs than the Raithel18 IFMR.
As compared to the fairly narrow mass distribution of BHs allowed by Raithel18 ($5 - 16 M_{\odot}$) the Spera15 IFMR allows for BHs as massive as 90 $M_{\odot}$ to form (Figure \ref{fig:IFMR_comp}). 

\subsection{The Sukhbold N20 IFMR}
\label{sec:Sukhbold N20 IFMR}

\subsubsection{$M_{ZAMS}$-$M_{BH}$ relationship}

\begin{figure}
    \includegraphics[width=\linewidth]{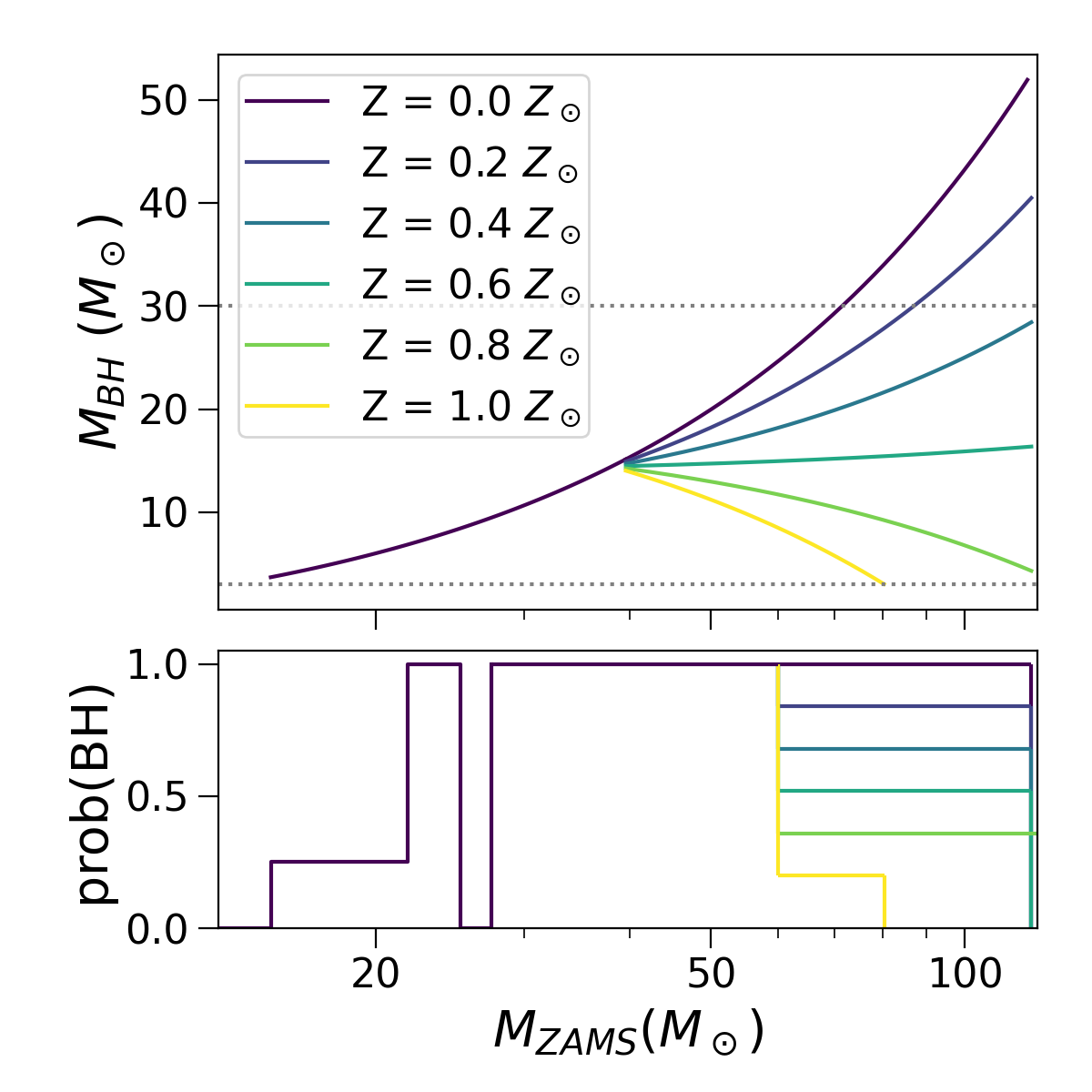}
\caption{\emph{Top:} The SukhboldN20 black hole initial-final mass relation at several different metallicities.
Below $M_{ZAMS} \lesssim 40M_\odot$, the BH remnant masses are identical independent of metallicity.
\emph{Bottom:} The probability of black hole formation at several different metallicities.
Below $M_{ZAMS} \lesssim 60M_\odot$, the probabilities are identical independent of metallicity.}
\label{fig:bh_ifmr}
\end{figure}

The SukhboldN20 IFMR is shown in Figure \ref{fig:bh_ifmr} and includes zero metallicity models from \citet{Sukhbold:2014}, solar metallicity models from the N20 set of \citet{Sukhbold:2016}, and pulsational-pair instability models from \citet{Woosley:2017, Woosley:2020}.
We collectively refer to these as the Sukhbold N20 simulations.

Any object with remnant mass $M_{rem} < 3 M_\odot$ is a NS, and those with $M_{rem} > 3 M_\odot$ are BHs.

\begin{deluxetable}{lccc}
\tablecaption{Compact Object Formation Probabilities 
    \label{tab:Compact Object Formation Probabilities}}
\tablehead{
    \colhead{Mass Range ($M_\odot$)} &
    \colhead{$P_{WD}$} & 
    \colhead{$P_{NS}$} & 
    \colhead{$P_{BH}$}
}
\startdata
0.5 $\leq M_{ZAMS} <$ 9.0 & 1.00 & 0.00 & 0.00 \\
9.0 $\leq M_{ZAMS} <$ 15.0 & 0.00 & 1.00 & 0.00 \\
15.0 $\leq M_{ZAMS} <$ 21.8 & 0.00 & 0.75 & 0.25 \\
21.8 $\leq M_{ZAMS} <$ 25.2 & 0.00 & 0.00 & 1.00 \\
25.2 $\leq M_{ZAMS} <$ 27.4 & 0.00 & 1.00 & 0.00 \\
27.4 $\leq M_{ZAMS} <$ 60.0 & 0.00 & 0.00 & 1.00 \\
60.0 $\leq M_{ZAMS} <$ $M_{up}$ & 0.00 & 0.8$f_Z$ & 1 - 0.8$f_Z$ \\
$M_{up}$ $\leq M_{ZAMS} \leq$ 120.0 & 0.00 & 1.00 & 0.00 \\
\enddata
\tablecomments{Probabilities of forming a white dwarf $P_{WD}$, neutron star $P_{NS}$, or black hole $P_{BH}$ as a function of ZAMS mass $M_{ZAMS}$ for the SukhboldN20 IFMR.
$f_Z$ is defined in Eq. \ref{eq:f_Z} and $M_{up}$ is defined in Eq. \ref{eq:mmax}.}
\end{deluxetable}

To obtain the BH IFMR, we use least squares minimization to find the best-fit line through BH masses $M_{BH}$ for BHs in the zero metallicity N20 models where $15 M_\odot < M_{ZAMS} < 70 M_\odot$. 
These best-fit lines are shown in Figure \ref{fig:bh_ifmr}, labeled as $Z = 0.0 Z_{\odot}$ and $Z = 1.0 Z_{\odot}$ respectively. 
The zero metallicity models are fit by
\begin{equation}
    \label{eq:zero_N20_BH_mass}
    M_{BH,0}(M_{ZAMS}) = 0.4652 M_{ZAMS} - 3.2917.
\end{equation}
Below $M_{ZAMS}\sim 40 M_\sun$, remnant mass is independent of metallicity, and Eq. \ref{eq:zero_N20_BH_mass} is applicable for all stars with $M_{ZAMS}\lesssim 40 M_\sun$.
However, above $M_{ZAMS}\sim 40 M_\sun$, the relationship between $M_{ZAMS}$ and $M_{rem}$ becomes metallicity dependent.
We fit a line using least squares minimization to the BH points between $40 M_\odot < M_{ZAMS} <70 M_\odot$ in the solar metallicity N20 models.
The solar metallicity N20 models above $M_{ZAMS} \sim 40 M_\sun$ are fit by
\begin{equation}
\label{eq:solar_N20_BH_mass}
    M_{BH,Z_\odot}(M_{ZAMS}) = -0.271 M_{ZAMS} + 24.743.
\end{equation}

To obtain the remnant mass as a function of metallicity for $M_{ZAMS} \gtrsim 40 M_\odot$, we linearly interpolate between the zero and solar metallicity models (Eqs. \ref{eq:zero_N20_BH_mass} and \ref{eq:solar_N20_BH_mass}).
The choice of linear interpolation is arbitrary, as the uncertainties in massive stellar evolution are so large the proper interpolation scheme is unknown.
However, the main trend is captured: for zero metallicity, the mass of the remnant black hole always increases, while for solar metallicity, mass loss eventually catches up for the high mass stars and the remnant mass decreases, with intermediate metallicities having behavior in between those two cases.

In addition, we include pulsational-pair instability supernovae (PPISN) by extrapolating Eqs. \ref{eq:zero_N20_BH_mass} and \ref{eq:solar_N20_BH_mass} out to $120 M_\odot$.
For high mass stars $100 M_\odot \lesssim M_{ZAMS} \lesssim 140 M_\odot$, electron-positron pair production robs the star of energy, causing it to eject a substantial fraction of its mass, resulting in a core-collapse SNe and remnant black holes of $\sim 35-50 M_\odot$.
The $M_{ZAMS}$ corresponding at which a star reaches the PPISN region increases with increasing metallicity.
At sufficiently high metallicities, there is no $M_{ZAMS}$ that reaches the PPISN region.

Putting all the above together, the N20 black hole IFMR is given by the following piecewise function:
\begin{equation}
\label{eq:MBH}
  M_{BH}(M_{ZAMS}, Z) = 
  \begin{cases}
    M_{BH,0}, \\ \quad 15 M_\odot< M_{ZAMS} < 39.6 M_\odot & \\
    (1 - f_Z) M_{BH,0} + f_Z M_{BH,Z_\odot}, \\ \quad 39.6 M_\odot < M_{ZAMS} < M_{up}
  \end{cases}
\end{equation}
where
\begin{equation}
  f_Z = 
  \begin{cases}
    Z/Z_\odot, & 0 < Z \leq Z_\odot \\
    1, & Z > Z_\odot
  \end{cases}
\label{eq:f_Z}
\end{equation}
and
\begin{equation}
\label{eq:mmax}
  M_{up} = 
  \begin{cases}
    120 M_\odot, &\quad f_Z \leq 0.63 \\
    \textrm{max}(120M_\odot, M'), &\quad f_Z > 0.63 
 \end{cases}
\end{equation}
where
\begin{equation}
    \label{eq:M'}
    M' = \frac{6.292 - 28.035 f_Z}{0.465 - 0.736 f_Z}  M_\odot.
\end{equation}
The condition on $M_{up}$ (Eqs. \ref{eq:mmax} and \ref{eq:M'}) ensures $M_{BH} \geq 3M_\odot$, while also restricting the maximum $M_{ZAMS}$ to $120M_\odot$. 
We also assume super-solar metallicity stars have the same behavior as solar metallicity stars (Eq. \ref{eq:f_Z}).

Note also that the BH masses from Eq. \ref{eq:MBH} are lower limits on the BH masses from the simulation, as they correspond to the He core mass of the star at the time of implosion.
It is uncertain whether the envelope is entirely ejected or whether some of it falls back onto the core and contributes to the black hole mass.
The difference between the He core mass and the total pre-supernova mass provide the limits of the remnant black hole mass \citep{Sukhbold:2016, Raithel:2018}.
Compared to observational data of BHs, \citet{Raithel:2018} found a high ejection fraction of the envelope and the remnant BH mass was quite similar to the He core mass for solar metallicity stars; however, it is not known whether this result is metallicity dependent. 
 
\subsubsection{$M_{ZAMS}$ vs. NS/BH formation probability}

Simulations indicate there is no $M_{ZAMS}$ above which BHs are always formed and below which NSs are formed \citep{Sukhbold:2016}.
Other factors such as the metallicity of the star or core structure directly prior to explosion also determine what type of remnant is left behind.
To include this stochasticity in the SukhboldN20 IFMR, we assign the different outcomes probabilistically based on $M_{ZAMS}$. 
The probabilities for NS vs. BH formation are taken from the Sukhbold N20 simulations.
We follow the approach of \citet{Raithel:2018}, by choosing the fewest number of bins possible to capture the different probability regions where no/some/only BHs are formed.
Below $M_{ZAMS}\sim 60 M_\sun$, the probability of NS or BH formation are independent of metallicity.
Similarly to the $M_{BH}-M_{ZAMS}$ relationship, to determine probability of a BH or NS remnant above $M_{ZAMS}\sim 60 M_\sun$, we interpolate linearly between the zero and solar metallicity models (Table \ref{tab:Compact Object Formation Probabilities}).
For $M_{up} \leq M_{ZAMS} \leq 120M_\odot$, the probability of BH formation is zero.
Although we take the neutron star formation probabilities from the Sukhbold N20 simulations, we do not use the masses.
Instead, we draw from the same neutron star mass distribution as for the Raithel18 IFMR (Appendix \ref{sec:Neutron Star Mass Distribution}). 

\begin{figure}
    \centering
    \includegraphics[scale=0.5]{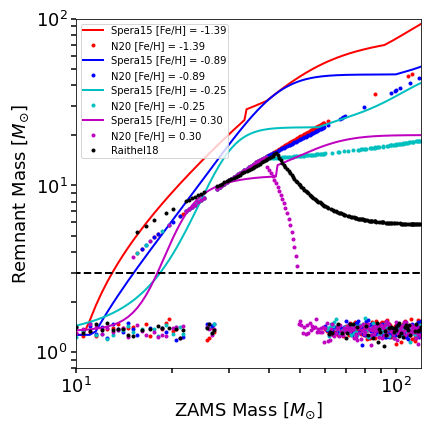}
    \caption{\edit{Remnant mass versus ZAMS mass for different IFMRs.
    The Raithel18 IFMR is defined for solar metallicity only.
    The Spera15 and SukhboldN20 IFMRs are a function of metallicity; in this figure they are evaluated at four different metallicities corresponding to the values described in \S \ref{sec:Metallicity Binning}.}
    The dashed black line at 3 $M_{\odot}$ represents the BH/NS boundary. 
    While the Spera15 IFMR is analytic, the SukhboldN20 and Raithel18 IFMRs are stochastic. 
    Even stars with high ZAMS masses can form NSs instead of BHs. 
    The differences in the IFMRs are most apparent at the highest stellar and compact object masses.}
    \label{fig:IFMR_comp}
\end{figure}

\begin{deluxetable*}{l|rr|rr|rrrr}
\tablecaption{\texttt{PopSyCLE} vs. \citet{Mroz:2019} Event Rates}
    \label{tab:PopSyCLE Fields}
\tablehead{
  \colhead{Name} & 
  \colhead{$l$} &
  \colhead{$b$} &
  \multicolumn{2}{c}{$n_{*}$ ($10^{6}$)} &
  \multicolumn{4}{c}{$\Gamma$ ($10^{-6}$)} \\
  \colhead{} & 
  \colhead{(deg)} &
  \colhead{(deg)} &
  \multicolumn{2}{c}{(stars deg$^{-2}$)} &
  \multicolumn{4}{c}{(events star$^{-1}$ yr$^{-1}$)} \\
  \colhead{} &
  \colhead{} &
  \colhead{} &
  \colhead{\citetalias{Mroz:2019}} &
  \colhead{Sim.} &
  \colhead{\citetalias{Mroz:2019}} &
  \colhead{Spera15} &
  \colhead{Raithel18}  &
  \colhead{SukhboldN20} \\
  \colhead{} &
  \colhead{} &
  \colhead{} &
  \colhead{} &
  \colhead{} &
  \colhead{} &
  \colhead{(Sim.)} &
  \colhead{(Sim.)} &
  \colhead{(Sim.)} }
\startdata
OGLE-IV-BLG500 & 1.00 & -1.03 & 4.84 & 3.37 & 23.9 $\pm$ 2.0 & 35.7 $\pm$ 3.4 & 39.5 $\pm$ 3.6 & 31.6 $\pm$ 3.2 \\ 
OGLE-IV-BLG504 & 2.15 & -1.77 & 8.47 & 3.10 & 16.9 $\pm$ 1.2 & 22.2 $\pm$ 2.8 & 20.8 $\pm$ 2.7 & 16.7 $\pm$ 2.4 \\ 
OGLE-IV-BLG506 & 0.01 & -3.00 & 9.19 & 3.83 & 16.5 $\pm$ 1.1 & 22.5 $\pm$ 2.5 & 21.0 $\pm$ 2.4 & 18.0 $\pm$ 2.2 \\ 
OGLE-IV-BLG511 & 3.28 & -2.52 & 9.61 & 3.64 & 13.5 $\pm$ 1.0 & 18.0 $\pm$ 2.3 & 15.9 $\pm$ 2.2 & 18.0 $\pm$ 2.3 \\ 
OGLE-IV-BLG527 & 8.81 & -3.64 & 4.54 & 2.04 & 5.5 $\pm$ 0.9 & 5.3 $\pm$ 1.7 & 7.4 $\pm$ 2.0 & 4.2 $\pm$ 1.5 \\ 
OGLE-IV-BLG611 & 0.33 & 2.82 & 4.95 & 3.66 & 16.2 $\pm$ 1.3 & 17.9 $\pm$ 2.3 & 18.8 $\pm$ 2.3 & 19.1 $\pm$ 2.4 \\ 
OGLE-IV-BLG629 & 7.81 & 4.81 & 3.26 & 1.49 & 3.4 $\pm$ 1.1 & 2.9 $\pm$ 1.4 & 2.9 $\pm$ 1.4 & 2.9 $\pm$ 1.4 \\ 
OGLE-IV-BLG648 & 1.96 & 0.94 & 2.04 & 1.24 & 18.3 $\pm$ 2.4 & 12.1 $\pm$ 3.2 & 8.6 $\pm$ 2.7 & 7.8 $\pm$ 2.6 \\ 
OGLE-IV-BLG675 & 0.78 & 1.69 & 4.03 & 3.94 & 26.5 $\pm$ 2.3 & 22.1 $\pm$ 2.5 & 28.6 $\pm$ 2.8 & 22.6 $\pm$ 2.5 \\ 

\enddata
\tablecomments{\edit{Observed vs. simulated stellar density and efficiency-corrected event rates for nine select fields in the OGLE-IV survey. 
To calculate the efficiency-corrected event rates for the Spera15, Raithel18, and SukhboldN20 IFMR simulations, we apply the following cuts to match the completeness-corrected sample in \citetalias{Mroz:2019}: source magnitude $I < 21$ mag, maximum impact parameter $u_0 < 1$, and Einstein crossing time range $0.5 < t_E < 300$ days).
The stellar densities for each field are also based on OGLE observability cuts ($I < 21$ mag), neglecting the effects of crowding.
Stellar rather than compact object events dominate the microlensing event rate; thus, changing the IFMR only has a small impact on the overall event rates, generally within the uncertainties.}}
\end{deluxetable*}

\section{Results}

\edit{Using the updated version of \texttt{PopSyCLE},} we simulated a total of nine 0.34 deg$^2$ fields, with each field centered at the location of an OGLE-IV Bulge field (Table \ref{tab:PopSyCLE Fields}).
\edit{Each simulation was 1000 days long, with a sampling cadence that detected all events with $t_E \gtrsim 3$ days.\footnote{\edit{Specifically, we ran the simulation with sampling cadence of 10 days.
Note that in \texttt{PopSyCLE}, the sampling cadence is \emph{not} equivalent to a real survey's observational cadence. 
This is why a sampling cadence of 10 days can detect events with $t_E < 10$ days.
See \S 4.3 and Figure 3 of \citet{Lam:2020} for further details.}}}
The chosen fields span Galactic longitudes 0$^\circ$ to 8$^\circ$ and Galactic latitudes -4$^\circ$ to 5$^\circ$.

\label{sec:Results}
\subsection{Impact of the IFMR on Compact Object Population}
\label{sec:Impact of the IFMR on Compact Object Populations}
\begin{figure*}
    \centering
    \includegraphics[scale=0.6]{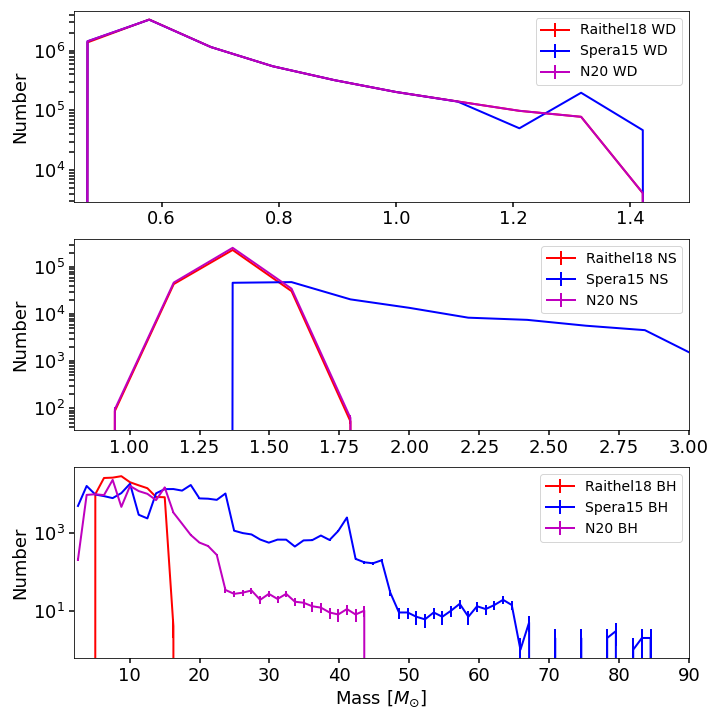}
    \caption{Mass distributions of all compact objects (WD, NS, BH) produced by each IFMR.
    \edit{The population synthesis was performed over an area of 0.34 deg$^2$ centered on the coordinates of the OGLE-IV-BLG611 field.
    The histograms show the underlying population of compact objects, i.e. lensing is not considered.}
    The WD mass distributions are nearly identical for all three IFMRs except at the high mass end where Spera15 has an excess of high mass WDs as compared to SukhboldN20 and Raithel18. 
    The NS mass distributions are similar for SukhboldN20 and Raithel18, while Spera15 has significantly more high mass NSs. 
    The BH mass distribution varies greatly depending on the IFMR.}
    \label{fig:COs_611}
\end{figure*}

The different IFMRs presented in this paper result in different underlying mass distributions for the Milky Way compact object populations. 
\edit{Figure \ref{fig:COs_611} compares the compact object populations produced by the different IFMRs in the simulated OGLE-IV-BLG611 field.} 

The WD population does not change significantly since all the IFMRs for low-mass stars are based on \citet{Kalirai:2008}.
\edit{The average mass of a WD in field OGLE-IV-BLG611 is 0.65 $M_{\odot}$, 0.66 $M_{\odot}$, and 0.64 $M_{\odot}$, for the Raithel18, Spera15, and SukhboldN20 IFMRs, respectively.}

\edit{The NS population produced by the Spera15 IFMR tends to be about 30\% more massive than those produced by the Raithel18 and SukhboldN20 IFMRs.
Although many neutron stars have masses below 1.4 $M_{\odot}$ (Appendix \ref{sec:Neutron Star Mass Distribution}), the Spera15 IFMR labels all compact objects below 1.4 $M_{\odot}$ as WDs, only allowing NSs in the $1.4 - 3 M_\odot$ mass range (\S \ref{sec:The Spera15 IFMR Function})
This causes the NSs from the Spera15 IFMR to be more massive on average than the NS populations of Raithel18 and SukhboldN20, both of which have their masses drawn from a Gaussian distribution with a mean of 1.36 $M_{\odot}$ (Appendix \ref{sec:Neutron Star Mass Distribution}).
The average mass of a NS in field OGLE-IV-BLG611 is 1.36 $M_{\odot}$, 1.75 $M_{\odot}$, and 1.36 $M_{\odot}$, for the Raithel18, Spera15, and SukhboldN20 IFMRs, respectively.} 

The most significant differences between the IFMRs are found in the BH mass distribution. 
\edit{The average mass of a BH in field OGLE-IV-BLG611 is 9.32 $M_{\odot}$, 14.74 $M_{\odot}$, and 9.99 $M_{\odot}$, for the Raithel18, Spera15, and SukhboldN20 IFMRs, respectively.}
The similarity in average BH mass between the Raithel18 and SukhboldN20 IFMR is indicative of the fact that most 
\edit{stars in the line of sight toward the Bulge are solar or super-solar metallicity, and the Raithel18 and SukhboldN20 IFMRs are similar in this metallicity regime (Figure \ref{fig:IFMR_comp}).
In addition, the Spera15 IFMR tends to produce much more massive BHs than either the Raithel18 or SukhboldN20 IFMRs, which is why its average BH mass is higher (Figure \ref{fig:IFMR_comp}).}

\subsection{\texttt{PopSyCLE} vs. OGLE Observed Event Rates}
\label{sec:Comparison against observations}

\edit{
In order to validate the results from \texttt{PopSyCLE}, we compare the stellar density and event rates for the OGLE-like simulated survey to the efficiency-corrected results presented in Table 7 of \citet{Mroz:2019}.
To replicate the observing conditions of the OGLE survey, in the \texttt{PopSyCLE} simulation we use a seeing-limited blending radius of $0.65''$ and make observations in the $I$-band filter.
For the stellar density comparison, our star count is restricted to stars with $I < 21$ to match \citet{Mroz:2019}.
For the efficiency-corrected event rate comparisons, we restrict the events in the simulation to have Einstein crossing times $0.5 < t_E < 300$ days, source magnitude $I_{src} < 21$, and impact parameter $u_0 < 1$ in order to match the completeness-corrected sample of \citet{Mroz:2019} (see also Table 4, column Mock Mr\'{o}z19, in \citet{Lam:2020}).}
\textcolor{black}{In Table \ref{tab:PopSyCLE Fields}, the simulated stellar densities and event rates produced by each IFMR are compared to the observed event rates and stellar densities of \citet{Mroz:2019}.}

\edit{
Across different IFMRs, the simulation event rates are comparable as stellar lensing events dominate over compact object lensing. 
Overall the simulated event rates compare well the the completeness-corrected event rates observed by \citet{Mroz:2019}. 
}

\edit{
While the event rates from the simulation are in reasonable agreement with the observed rates, the stellar density of each field as reported in \citet{Mroz:2019} is typically a factor of 2 higher than \texttt{PopSyCLE} predicts. 
This is a known issue found in other Galactic models, likely due to uncertainties in the length, angle, and overall structure of the Galactic bar, as well as variable extinction over small scales toward the Bulge.
The factor of 2 difference in star counts cannot be explained by accounting for stellar binarity or confusion (Abrams et al. in preparation).
Additional details are presented in Appendix \ref{sec:Galactic Model Comparisons}.}

\subsection{BH Microlensing Statistics with OGLE}
\label{sec:BH Microlensing Statistics}

\edit{We next consider whether differences in the IFMR are detectable from the observed (i.e. non-completeness corrected) distributions of microlensing events.
To do this, we exclude simulated \texttt{PopSyCLE} microlensing events that are faint (baseline magnitude $I_{base} > 21$), not substantially lensed (impact parameter $u_0 > 2$), or have a low observed amplification $\Delta m < 0.1$ mag.
These cuts were based on the events reported by OGLE's Early Warning System \citep[EWS,][]{Udalski:1994} from 2016-2018.
See also Table 4, column Mock EWS, in \citet{Lam:2020}.}

\begin{figure}
    \centering
    \includegraphics[scale=0.5]{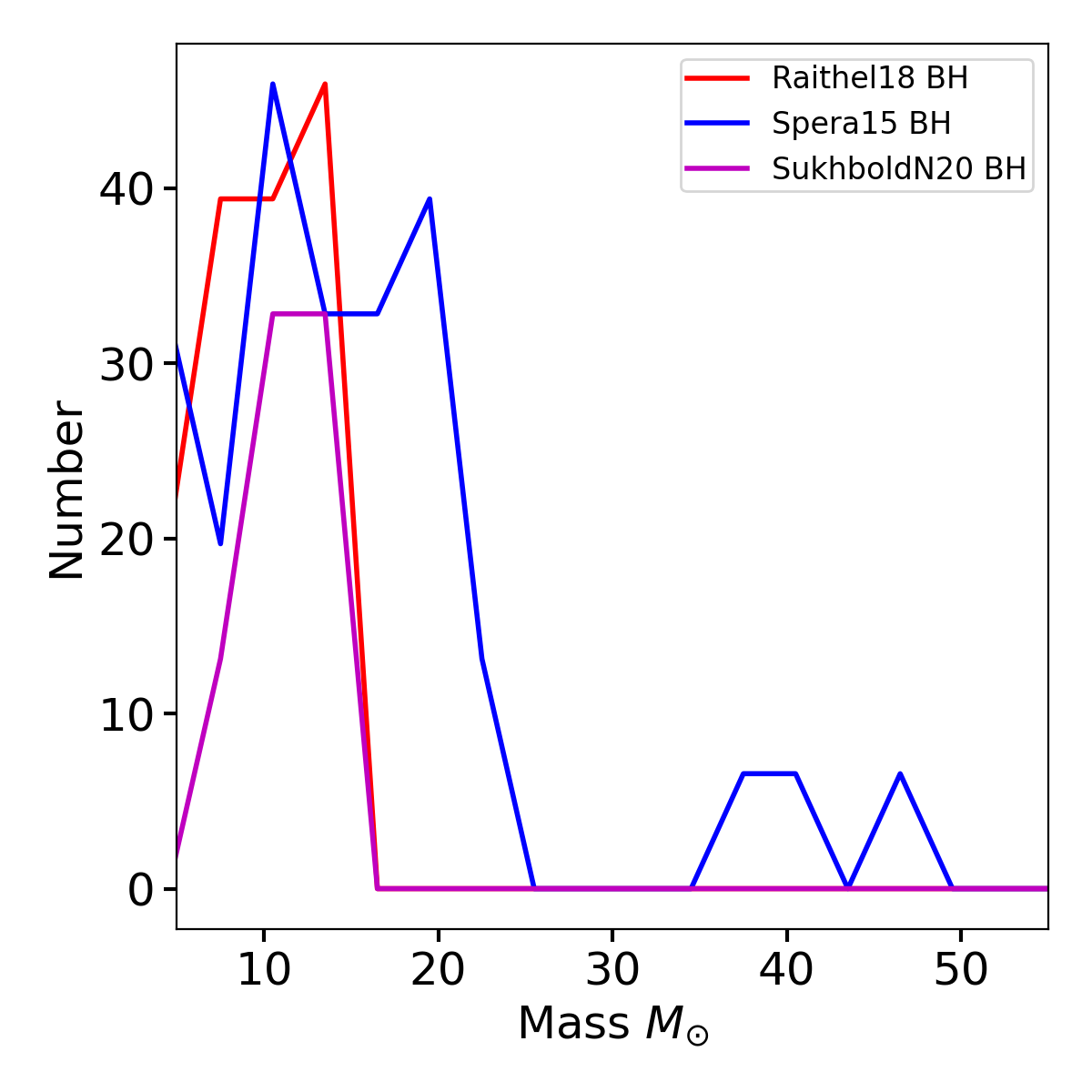}
    \caption{\edit{Mass distribution of black hole lenses from OGLE-detectable microlensing events.
    The number of events is scaled to the number of events an OGLE-like survey would observe over a span of 10 years, as described in \S \ref{sec:Constraining the IFMR with OGLE}.
    The Spera15 IFMR produces much more massive BHs than the Raithel18 and SukhboldN20 IFMRs, and this is reflected in the detected lens masses. 
    The SukhboldN20 IFMR produces the fewest BH lensing events overall}.}
    \label{fig:BH_lens_mass_fig}
\end{figure}

\edit{The mass distribution of black hole lenses for the different IFMRs as detectable by a 10 year OGLE Galactic Bulge survey is shown in Figure \ref{fig:BH_lens_mass_fig}.}
The Spera15 IFMR has significantly more high-mass lenses compared to either Raithel18 or SukhboldN20 IFMRs. 

\edit{To explore whether these mass distribution differences are detectable, we investigate their impact on observable quantities measured by photometric microlensing surveys.}
All microlensing parameters depend on the Einstein radius $\theta_E$ 
\begin{equation}
\label{eq:theta_E}
    \theta_E = \sqrt{\frac{4GM_L}{c^2} \Bigg( \frac{1}{d_L} - \frac{1}{d_S} \Bigg)}
\end{equation}
with $d_{L}$ the distance to the lens, $d_{S}$ the distance to the source, and $M_{L}$ the mass of the lens.
However, $\theta_E$ is generally not measurable with a photometric microlensing lightcurve; only quantities normalized by $\theta_E$ can be measured.
This includes the Einstein crossing time $t_E$ and the microlensing parallax $\pi_E$.

The Einstein crossing time is
\begin{equation}
\label{eq:t_E}
    t_E = \frac{\theta_E}{\mu_{rel}}
\end{equation}
where $\mu_{rel}$ is the magnitude of the relative source-lens proper motion.
The Einstein crossing time characterizes the length of the photometric microlensing event. 

The microlensing parallax is
\begin{equation}
\label{eq:pi_E}
    \pi_E = \frac{\pi_{rel}}{\theta_E}
\end{equation}
where $\pi_{rel}$ is the relative parallax
\begin{equation}
\label{eq:pi_rel}
    \pi_{rel} = 1 \mathrm{AU} \bigg( \frac{1}{d_L} - \frac{1}{d_S} \bigg)
\end{equation}
Microlensing parallax characterizes \edit{changes to the shape of the otherwise symmetric} photometric light curve due to the Earth's motion around the Sun.
It encodes information about the relative distance between the source and the lens. 

The Einstein crossing time and microlensing parallax scale with the lens mass as
\begin{equation}
\label{eq:t_E_scale}
    t_E \propto \sqrt{M_{L}}
\end{equation}
and
\begin{equation}
\label{eq:pie_E_scale}
    \pi_{E} \propto 1/\sqrt{M_{L}}.
\end{equation} 
As shown by \citet{Lam:2020} the most massive lenses (i.e. BHs) are characterized by a \edit{long} Einstein crossing time and a \edit{small} microlensing parallax. 

\begin{figure*}
    \centering
    \includegraphics[scale=0.6]{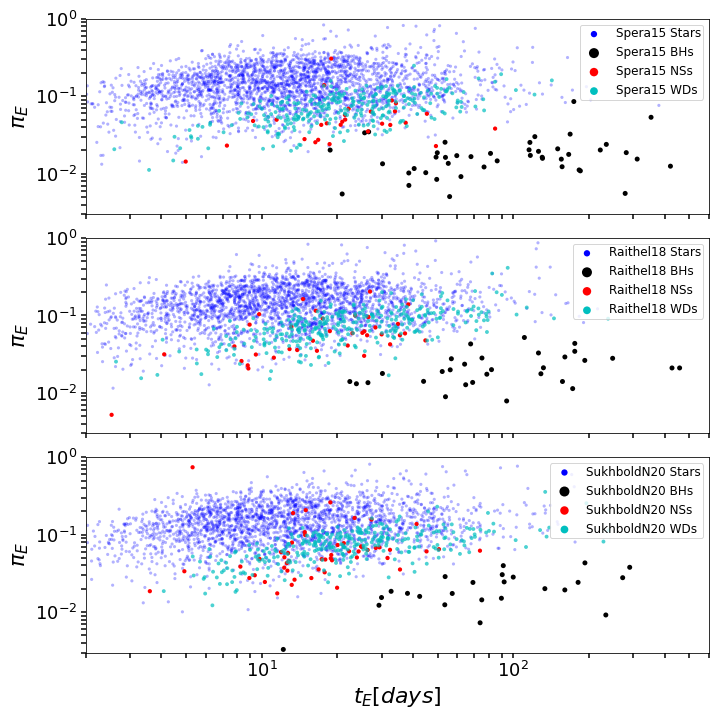}
    \caption{Microlensing parallax versus Einstein crossing time for \edit{OGLE-detectable microlensing events (described in the first paragraph of \S \ref{sec:BH Microlensing Statistics}) as a function of} lens type. 
    Regardless of IFMR, black hole lensing events are characterized by their long $t_E$ and low $\pi_E$ as compared to other types of lenses. The addition of the more massive Spera15 BH lenses allows for even smaller microlensing parallaxes indicating that events with very low $\pi_E$ even at shorter $t_E$ are likely to be good black hole microlensing candidates.}
    \label{fig:piE_tE_plot}
\end{figure*}

\edit{The Spera15 IFMR allows a wider range of black hole masses, and the more massive black holes produce longer Einstein crossing times and smaller microlensing parallaxes as compared to the SukhboldN20 and Raithel18 IFMRs as shown in Figure \ref{fig:piE_tE_plot}. 
The Einstein crossing time is not as sensitive to changes in the lens mass as the microlensing parallax, since the relative proper motion also affects the Einstein crossing time.}
For these reasons, the Spera15 BH lens population in  $\pi_{E}$ versus $t_{E}$ space does not change dramatically in $t_{E}$, but includes events with much lower microlensing parallax as compared to the Raithel18 BH lensing population. 
Because the SukhboldN20 BH lens population does not have a significantly different mass distribution as compared to the Raithel18 BH lens population, it is difficult to distinguish them based on their resultant distributions of $\pi_{E}$ and $t_{E}$. 
This reflects the fact that there are far fewer low metallicity massive stars in the Milky Way then there are solar or super-solar metallicity ones.
In the solar and super-solar metallicity regimes, the SukhboldN20 and Raithel18 IFMRs are similar (Figure \ref{fig:IFMR_comp}). 

\begin{figure*}
    \centering
    \includegraphics[scale=0.6]{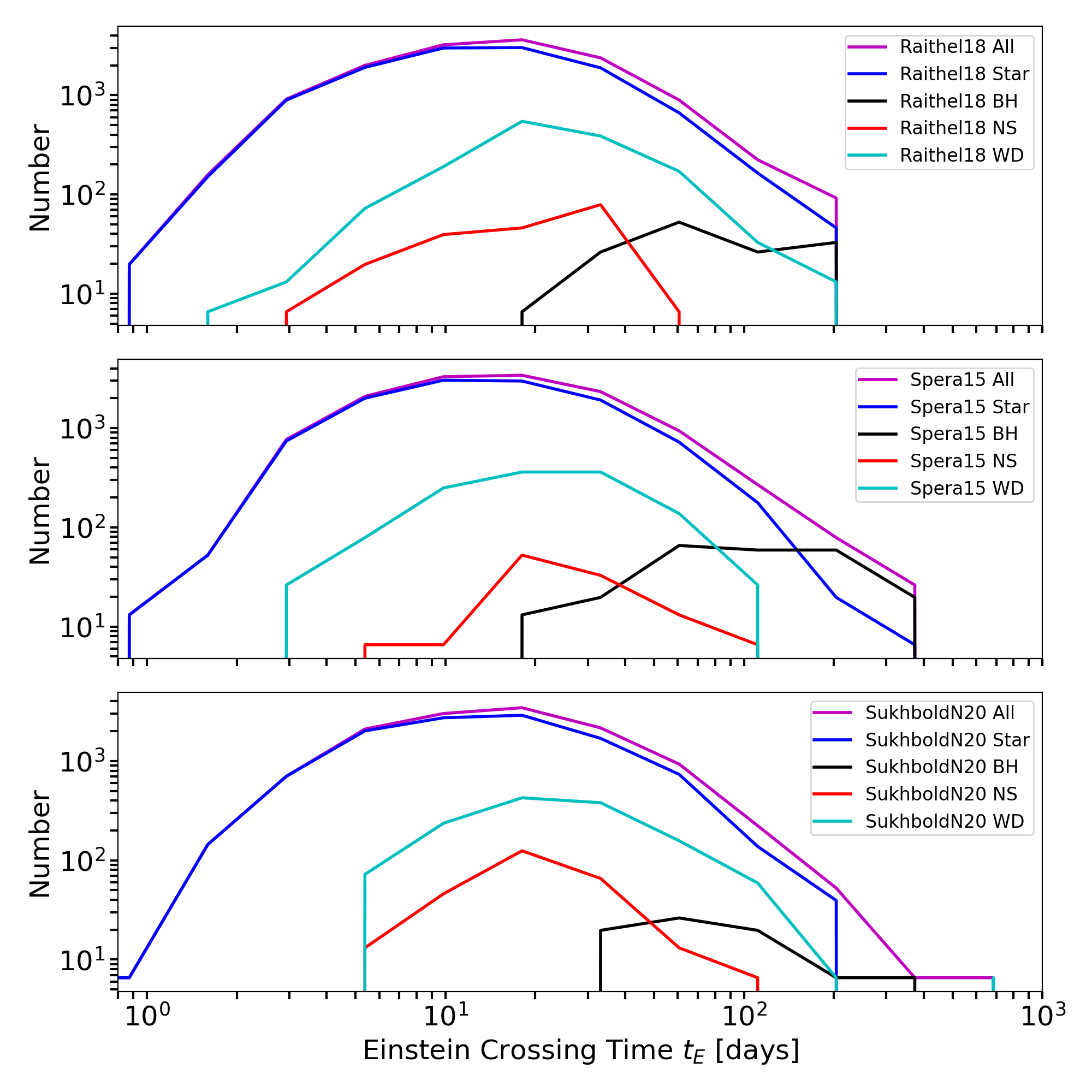}
    \caption{A histogram of Einstein crossing times for \edit{OGLE-detectable microlensing events as a function of} lens type for each IFMR. 
    \edit{The number of events is scaled to the number of events an OGLE-like survey would observe over a span of 10 years, as described in \S \ref{sec:Constraining the IFMR with OGLE}.}
    The largest differences between the IFMRs are found at \edit{long Einstein crossing times ($\gtrsim 150$ days)} as that is where the overall distribution is dominated by BH lenses.}
    \label{fig:tE_hist}
\end{figure*}

\begin{figure*}
    \centering
    \includegraphics[scale=0.6]{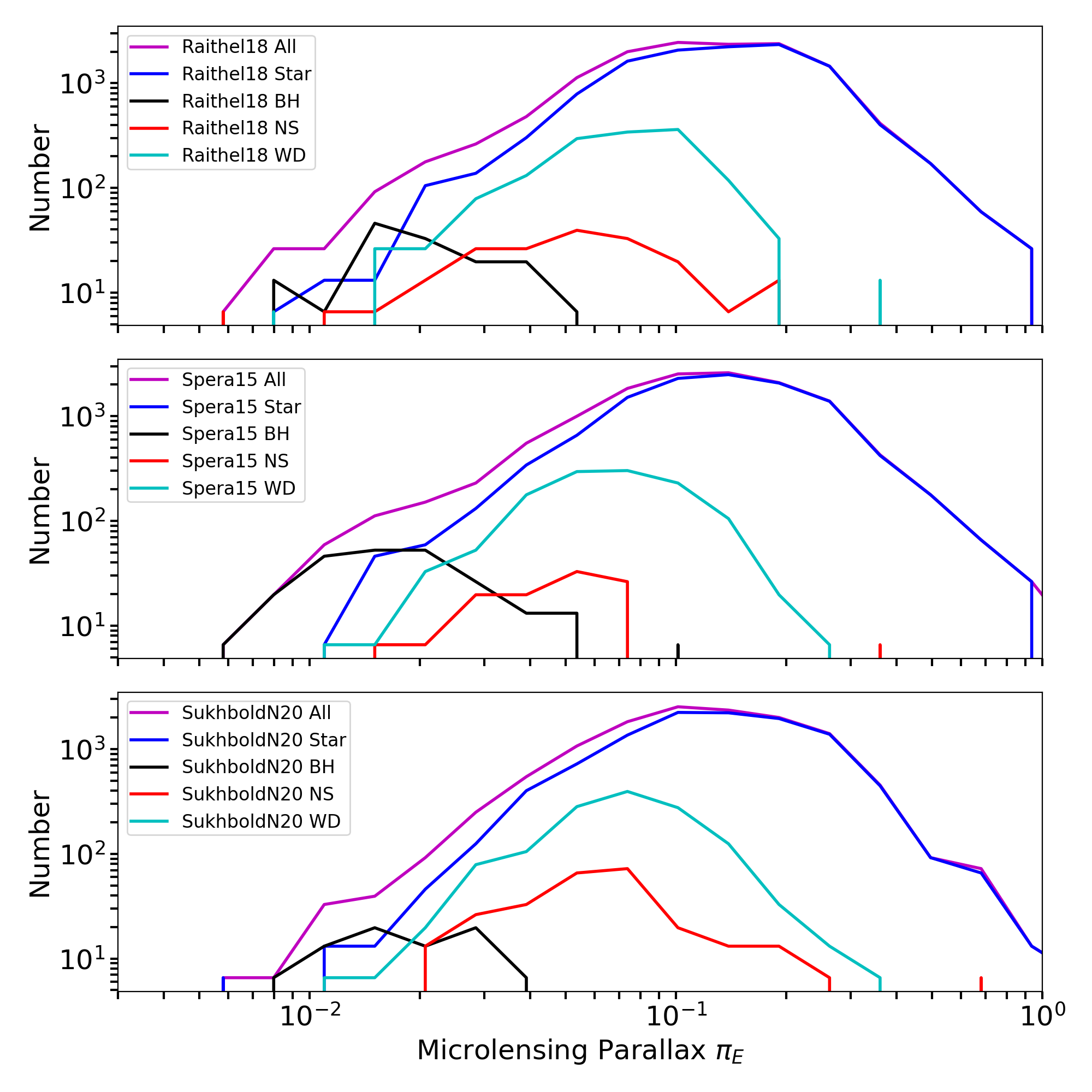}
    \caption{A histogram of microlensing parallax for \edit{OGLE-detectable microlensing events as a function of} lens type for each IFMR.    \edit{The number of events is scaled to the number of events an OGLE-like survey would observe over a span of 10 years, as described in \S \ref{sec:Constraining the IFMR with OGLE}.}
    The largest differences between the IFMRs are found at \edit{small microlensing parallaxes ($\lesssim 0.02$)} as that is where the overall distribution is dominated by BH lenses.}
    \label{fig:piE_hist}
\end{figure*}

\edit{An analysis of the distributions of $\pi_E$ and $t_E$ can be used to statistically constrain the Milky Way BH population \citep{Golovich:2022}.}
The difference in the compact object mass distribution between the IFMRs is reflected in the difference between the distributions for Einstein crossing time and microlensing parallax as shown in Figures \ref{fig:tE_hist} and \ref{fig:piE_hist}. 
The differences between the Spera15, Raithel18, and SukhboldN20 IFMRs are largest at low microlensing parallax. 
The effect of increasing lens mass on Einstein crossing time is weaker. 
For this reason the best way to find events that are likely to be caused by higher mass black holes is to select events not only with long Einstein crossing times, but also with very low microlensing parallax (\S \ref{sec:Defining the Milky Way BH Microlensing Sample} and \S \ref{sec:fisher matrix stuff}). 

\subsection{Constraining the IFMR with OGLE}
\label{sec:Constraining the IFMR with OGLE}

\edit{We now consider whether these different IFMRs are statistically distinguishable with 10 years of observations from an OGLE-like microlensing survey.
In this case, we are not interested in efficiency-corrected number, but rather the observed number of events.
Taking all the observable events as outlined in \S \ref{sec:BH Microlensing Statistics} and then rescaling the number to have the simulated area and duration match those of the OGLE survey would result in an overestimation of the observed number of events, as those observational cuts do not capture  sources of detection inefficiency such as observational gaps or sparse lightcurve coverage.
We thus empirically rescale the number of events we observe in our simulated subset of the OGLE survey to the expected number of total events as follows.}

\edit{\citet{Mroz:2017} published 2617 point-source point lens (PSPL) events in OGLE-IV's 9 high cadence Bulge fields from 2010-2015.
\citet{Mroz:2019} published 5790 PSPL events in the remaining 112 OGLE-IV low cadence Bulge fields from 2010-2017.
This corresponds to roughly 523 events/year and 827 events/year in the low and high cadence fields respectively, for a total of 1350 events/year total.
This implies over a 10 year survey, an OGLE-like survey should observe around 13500 events.
We thus scale the total number of observable simulated events produced by the Raithel18 IFMR to 13500.
Applying the same scaling factor to the SukhboldN20 and Spera15 IFMR simulations result in 12800 and 13300 events, respectively.}

\begin{figure}
    \centering
    \includegraphics[scale=0.5]{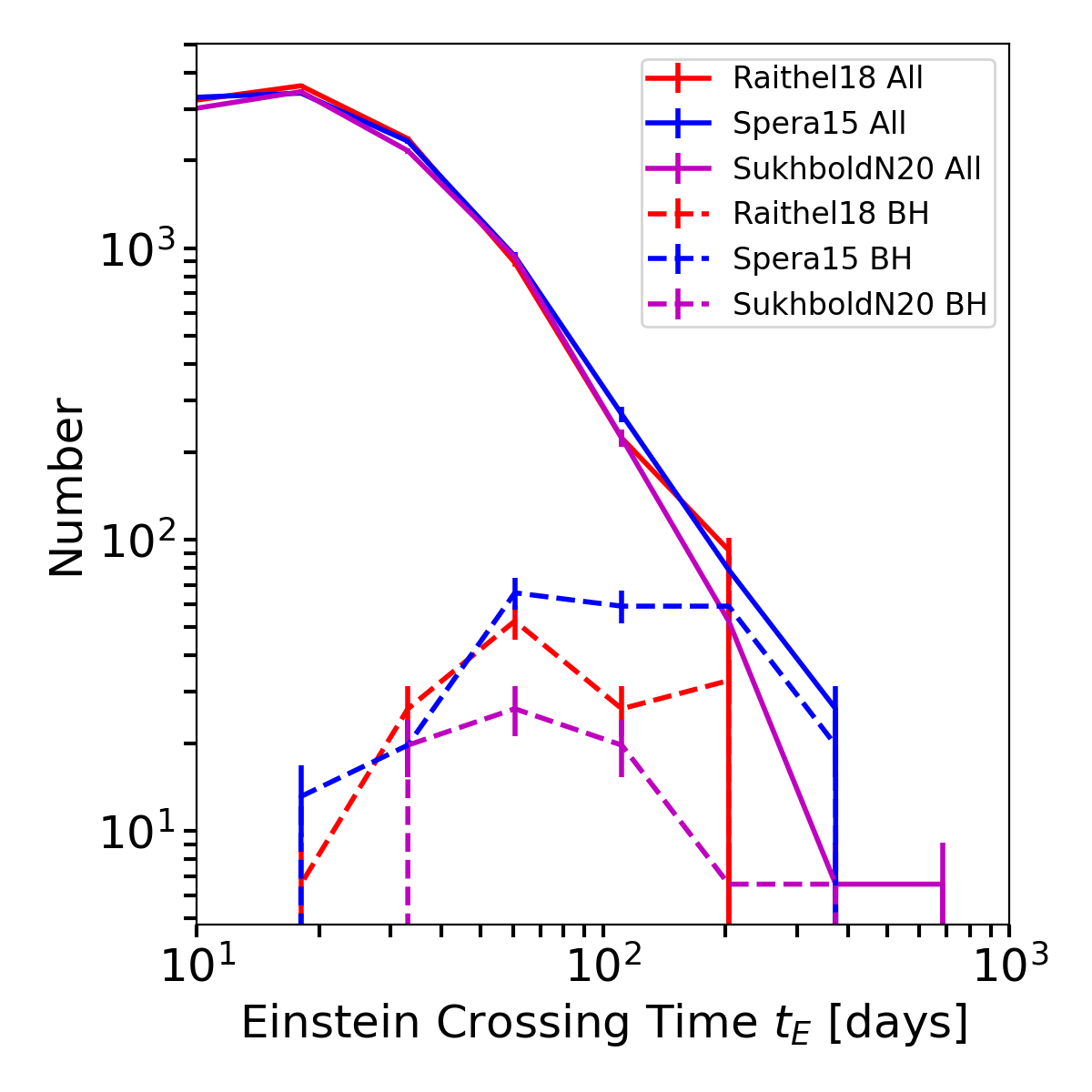}
    \caption{A histogram of Einstein crossing times for \edit{OGLE-detectable microlensing events.
    The number of events is scaled to the number of events an OGLE-like survey would observe over a span of 10 years, as described in \S \ref{sec:Constraining the IFMR with OGLE}.}
    The solid lines indicate all lens types while the dashed lines indicate the distribution of the black hole lensing events. 
    As $t_{E}$ $\propto$ $\sqrt{M_{L}}$ the slight excess of long duration events of the Spera15 IFMR can be attributed to the more massive BHs it produces as compared to the Raithel18 and SukhboldN20 IFMRs.}
    \label{fig:tE_OGLE}
\end{figure}

\begin{figure}
    \centering
    \includegraphics[scale=0.5]{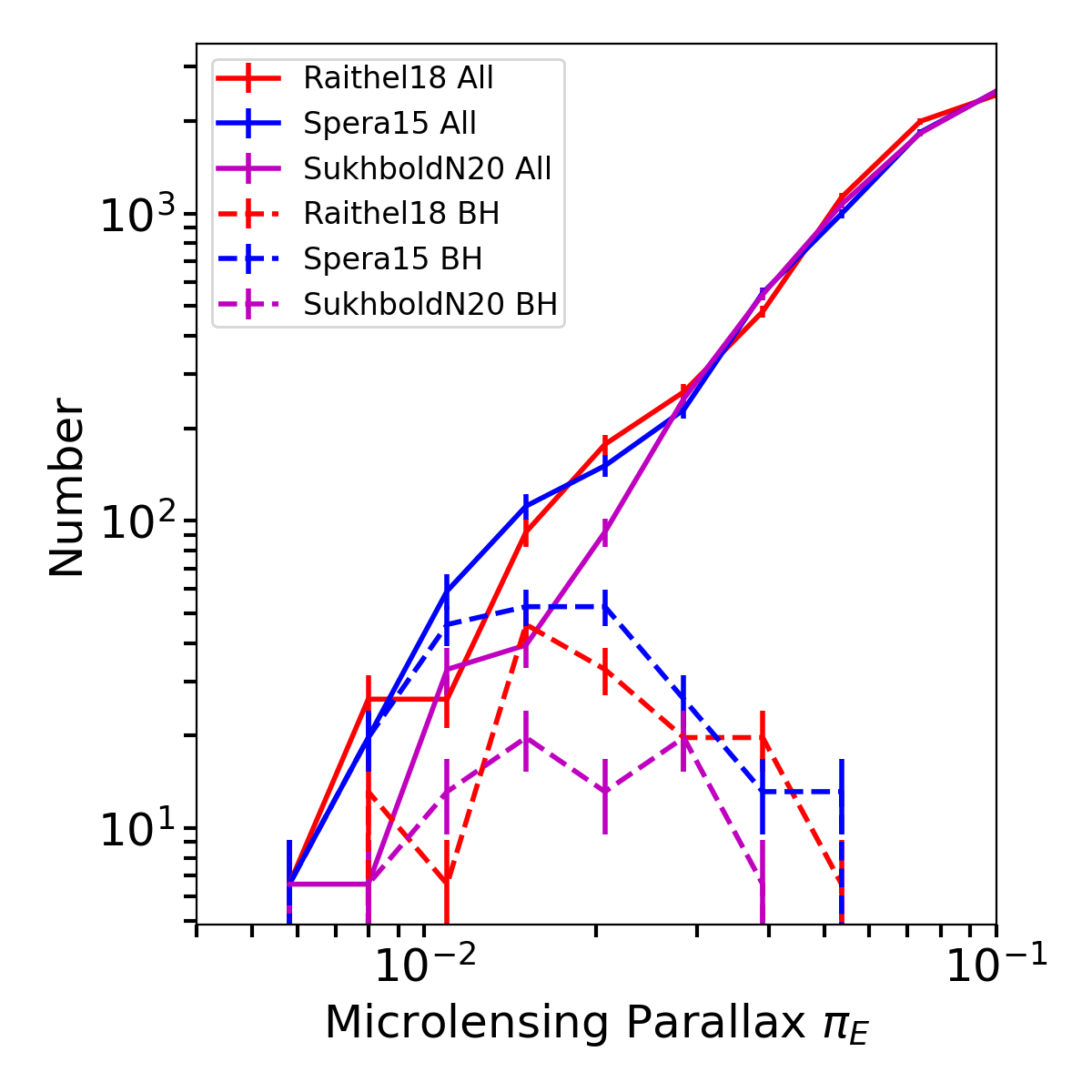}
    \caption{A histogram of microlensing parallax for 
    \edit{OGLE-detectable microlensing events.
    The number of events is scaled to the number of events an OGLE-like survey would observe over a span of 10 years, as described in \S \ref{sec:Constraining the IFMR with OGLE}.}
    The solid lines indicate all lens types while the dashed lines indicate the distribution of the black hole lensing events. 
    As $\pi_{E}$ $\propto$ $1/\sqrt{M_{L}}$ the excess of low $\pi_E$ events of the Spera15 IFMR can be attributed to the more massive BHs it produces as compared to the Raithel18 and SukhboldN20 IFMRs.}
    \label{fig:piE_OGLE}
\end{figure}

As discussed in \S \ref{sec:BH Microlensing Statistics} we expect the Spera15 IFMR to have an excess of high Einstein crossing time events and low microlensing parallax as compared to the Raithel18 and SukhboldN20 IFMRs, since the Spera15 IFMR \edit{produces more} high mass black holes. 
Figure \ref{fig:tE_OGLE} shows a slight excess of long Einstein crossing time events for the Spera15 IFMR as compared to the Raithel18 IFMR and the SukhboldN20 IFMR.
Figure \ref{fig:piE_OGLE} shows an excess of low microlensing parallax events for the Spera15 IFMR as compared to both the Raithel18 IFMR and the SukhboldN20 IFMR. 

\edit{Next, we focus on whether the Spera15 and SukhboldN20 IFMRs can be distinguished via $t_E$ and $\pi_E$.}
Integrating over the total Einstein crossing time distributions produced by each IFMR for events with $t_{E} > 100$ days we find a \edit{insignificant ($\lesssim 2 \sigma$) difference} in the number of events with Einstein crossing time greater than 100 days.
Integrating over the total microlensing parallax distributions to find the number of events with $\pi_{E} < 0.02$ produced by each IFMR, we find a \edit{significant $\sim5\sigma$} difference.
This indicates that low microlensing parallax is a more sensitive indicator of high lens mass (i.e. a black hole lens) than a long Einstein crossing time.

\edit{Unfortunately, these small microlensing parallaxes are not possible to measure with ground-based photometry.
Even for the long $t_E \approx 100 - 300$ days events where microlensing parallaxes are easier to constrain, OGLE is not very sensitive to $\pi_E \lesssim 0.03$ \citep[][Figure 3]{Wyrzykowski:2016}.}
For shorter duration events the precision with which microlensing parallax can measured is even worse. 
This means that most BH lensing events with their very small microlensing parallaxes will actually have undetectable parallaxes and are often not reported \citep{Karolinski:2020}.
\edit{Thus, with current survey capabilities, different IFMRs cannot be distinguished reliably using photometric microlensing.}

\textcolor{black}{Comparing the overall Einstein crossing time distribution reported in \citet{Mroz:2019} to the distributions produced by the \texttt{PopSyCLE} simulations (see Figure \ref{fig:tE_OGLE}), we note that there is a slight discrepancy in the peak of the distribution, with our models predicting a peak at around 20 days while the \citet{Mroz:2019} distribution peaks at around 25 days.
This difference is not unexpected as our simulation does not exactly reproduce the OGLE field of view, and also does not yet include the effect of binary lenses and sources which likely shift the peak of the Einstein crossing time distributions to longer timescales (Abrams et al., in prep).
Overall a 5 day shift of the Einstein crossing time distribution will not significantly effect the fraction of long duration events ($t_E > 100$ days) used to constrain the IFMR.
For a more extensive discussion of the comparisons of \texttt{PopSyCLE} simulations to OGLE observations please see \citet{Lam:2020}.
}

\subsection{Constraining the IFMR with the Roman Space Telescope}
\label{sec:Constraining the IFMR with the Roman Space Telescope}

The Nancy Grace Roman Space Telescope is an \edit{upcoming NASA flagship mission with a 2.4 meter telescope planned for launch in 2026-2027.
As one of its Core Community Surveys, Roman will conduct a microlensing survey toward the Galactic Bulge in the infrared to find thousands of cold exoplanets \citep{Penny:2019}.
In addition to photometric microlensing, Roman will also simultaneously be able perform astrometric microlensing measurements, allowing lens masses to be directly constrained.
\citet{Lam:2020} estimated that Roman's microlensing survey would be able to detect hundreds of black holes.}

\edit{We now consider whether Roman can distinguish different IFMRs solely with photometric measurements.
As an infrared space mission, Roman has advantages over an optical ground based survey like OGLE.}
Infrared observations allow for a larger number of sources to be detected as most sources will be low mass stars which emit at longer wavelengths.
Observations in H-band will also not suffer as much from interstellar extinction, which is especially important toward the Galactic Bulge. 
Roman will also be able to see sources with baseline magnitude less than 24th magnitude in H band, and as deep as 26th magnitude if the data is stacked \citep{Penny:2019} as compared to OGLE which is limited to sources with baseline magnitudes brighter then 22 magnitude in I band \citep{Udalski:1994}. 

Of the simulation fields in Table \ref{tab:PopSyCLE Fields}, field OGLE-IV-BLG500 is roughly located near the center of the proposed Roman exoplanet microlensing survey \citep{Penny:2019}. 
Fields OGLE-IV-BLG506 and OGLE-IV-BLG675 are also near the proposed Roman fields.
\edit{We thus use these three fields to represent the Roman sample used for this analysis.
To replicate the observing capabilities of Roman, in the \texttt{PopSyCLE} simulation we use a diffraction-limited blending radius of $0.09''$ and make observations in the H-band filter.
We define events detectable by Roman if they have baseline magnitude $I_{base} < 24$, impact parameter $u_0 < 2$, and observed amplification $\Delta m > 0.05$ mag (see also Table 4, column Mock WFIRST, in \citet{Lam:2020}.\footnote{Note that \citet{Lam:2020} allowed $I_{base} < 26$; we chose a brighter limit of $I_{base} < 24$ as a more conservative estimate that does not rely on stacking multiple observations together.
On the other hand, we allow for smaller amplitude microlensing events by requiring $\Delta m > 0.05$ mag, while  \citet{Lam:2020} required $\Delta m > 0.1$ mag; we assume Roman's photometric precision and sub-hour sampling rate will trim out low-amplitude variables better than OGLE.})
The number of microlensing events detected by the \texttt{PopSyCLE} simulation is then scaled to equal a 1.97 deg$^2$ survey area and 5 year duration to roughly match the number of events expected to be observed by Roman's microlensing survey. 
However, this method of determining Roman observable events neglects the effects of observational gaps, which lower the detection efficiency.}

\begin{figure}
    \centering
    \includegraphics[scale=0.5]{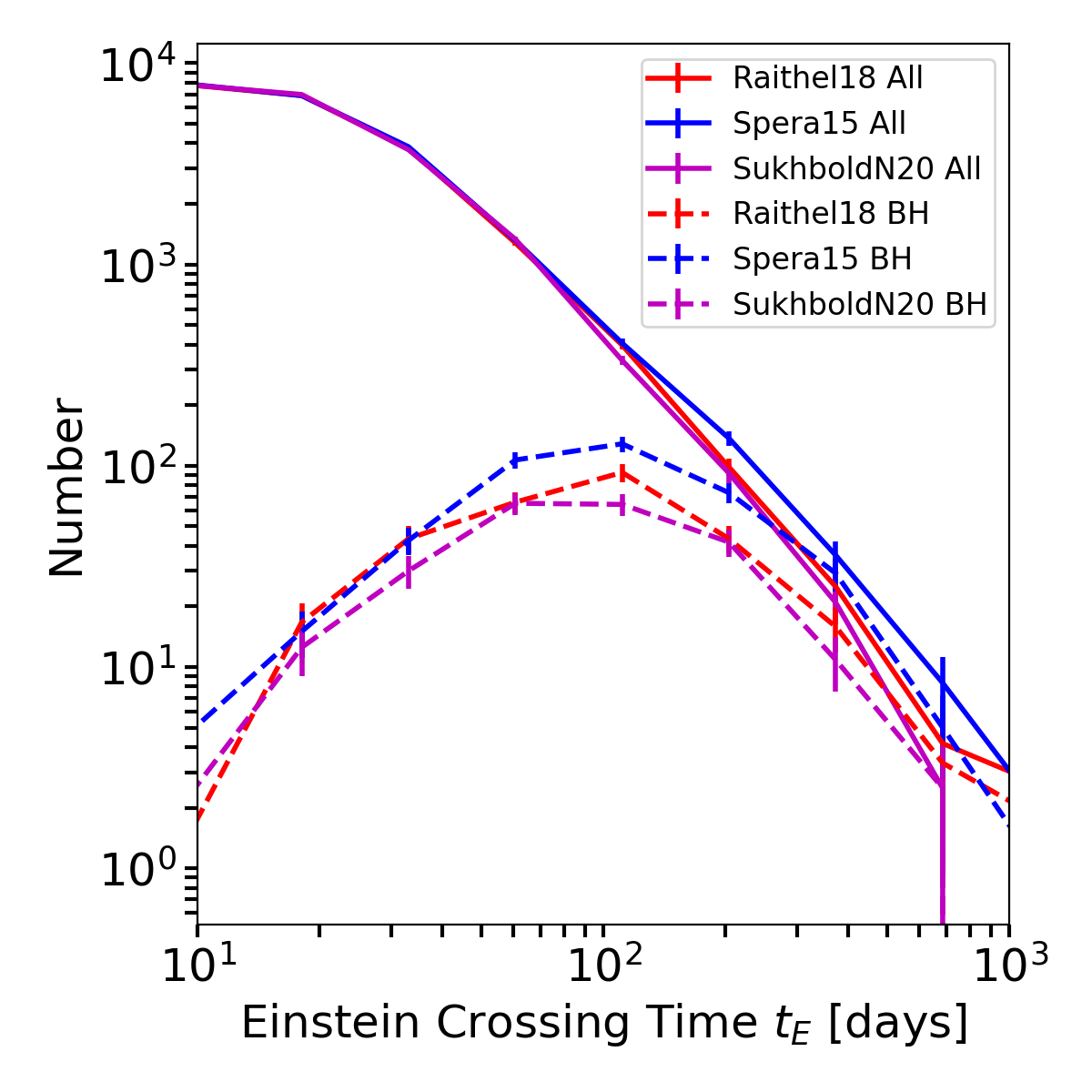}
    \caption{A histogram of Einstein crossing times for microlensing events detectable by a Roman style survey. \edit{The number of events is scaled to the number of events Roman is expected to detect during its 5 year microlensing survey, as described in \S \ref{sec:Constraining the IFMR with the Roman Space Telescope}.}
    The solid lines indicate all lens types while the dashed lines indicate the distribution of the black hole lensing events.}
    \label{fig:tE_Roman}
\end{figure}

\begin{figure}
    \centering
    \includegraphics[scale=0.5]{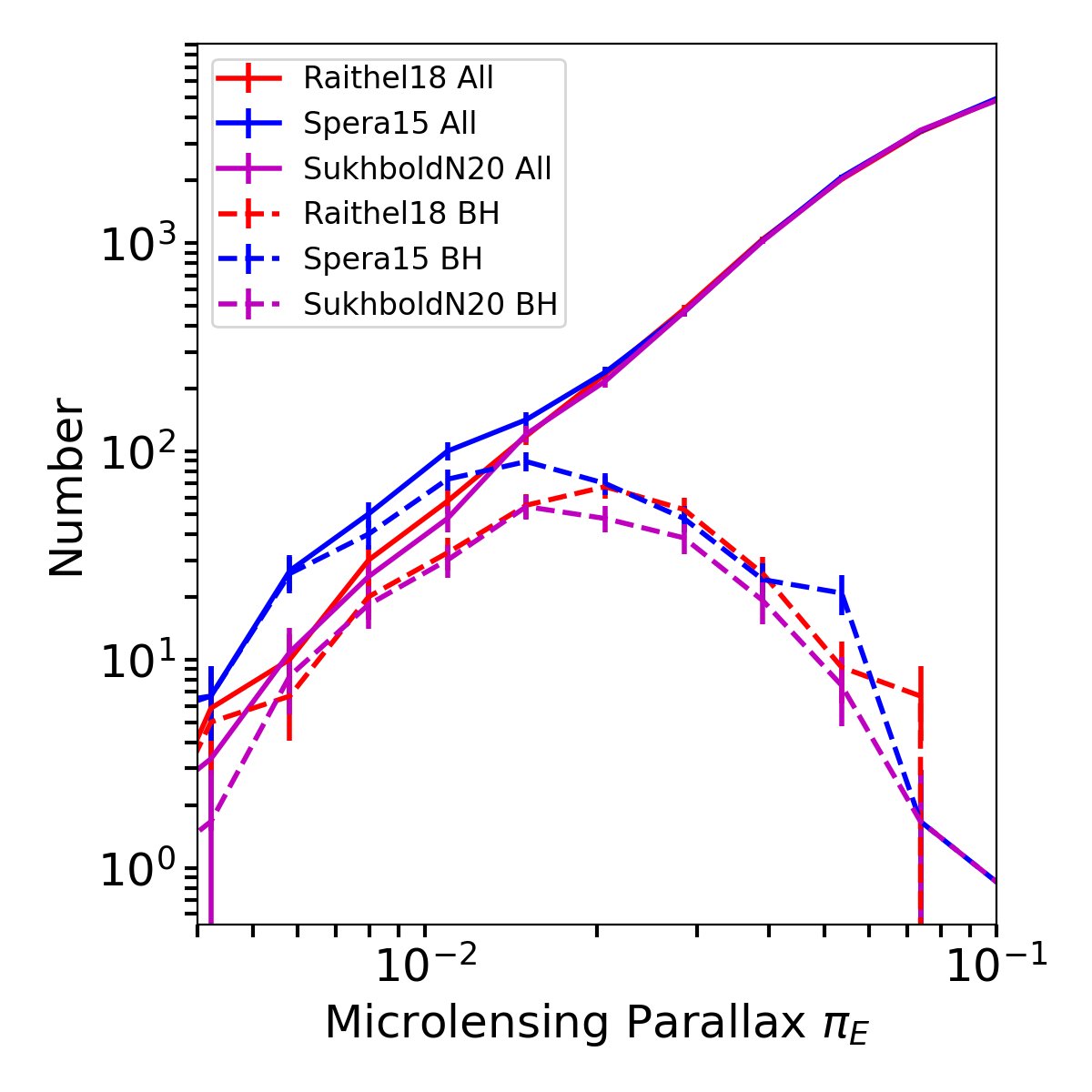}
    \caption{A histogram of microlensing parallaxes for microlensing events detectable by a Roman style survey. \edit{The number of events is scaled to the number of events Roman is expected to detect during its 5 year microlensing survey, as described in \S \ref{sec:Constraining the IFMR with the Roman Space Telescope}.}
    The solid lines indicate all lens types while the dashed lines indicate the distribution of the black hole lensing events.}
    \label{fig:piE_Roman}
\end{figure}

\edit{The fiducial Roman microlensing survey as presented in \citet{Penny:2019} consists of six 72-day Galactic Bulge observing seasons.
The seasons are centered on the vernal and autumnal equinoxes, with an observational gap of about 3.5 months between available observing windows; this is due to spacecraft pointing limitations due to the observatory design.
A complete analysis simulating Roman's detection efficiency is beyond the scope of this paper.
Instead, we make a simple estimate of the detection efficiency.
Based on $6 \times 72$ days of observations over a 5 year window, this corresponds to a duty cycle of 24\%.
We thus consider 24\% of the events detected by the criterion of the previous paragraph to be a sufficiently reliable estimate of the true number of microlensing events detectable by Roman.
This ultimately corresponds to 27,300, 27,300, and 27,500 microlensing events detected by Roman's 5 year microlensing survey for the Raithel18, SukhboldN20, and Spera15 IFMRs, respectively.
For comparison, Table 2 of \citet{Penny:2019} estimates 27,000 and 54,000 microlensing events with $|u_0| < 1$ and $|u_0| < 3$, respectively, to be detectable by Roman.
Their simulation, detection criteria, and method for estimating the observable number of events are quite different from ours; this demonstrates that our simple estimates are reasonable.
Thus we proceed with the assumption of 27,300-27,500 Roman-detectable microlensing events in the following results.}
Figures \ref{fig:tE_Roman} and \ref{fig:piE_Roman} show the expected distribution of Einstein crossing times and microlensing parallaxes detected by Roman assuming a Spera15 IFMR, SukhboldN20 or Raithel18 IFMR. 

\edit{Next, we focus on whether the Spera15 and SukhboldN20 IFMRs can be distinguished via $t_E$ and $\pi_E$.
Integrating over the total Einstein crossing time distributions produced by each IFMR for events with $t_{E} > 100$ days we find a significant ($\sim5 \sigma$) difference in the number of events with Einstein crossing time greater than 100 days.
Integrating over the total microlensing parallax distributions to find the total number of events with $\pi_{E} < 0.02$ produced by each IFMR, we also find a significant $\sim5\sigma$ difference.}

\edit{Although it will have exquisite photometric precision, it is uncertain if Roman will be able to measure such small microlensing parallaxes accurately due to the observational gaps. 
In future work, we will explore how additional observations during Roman's Bulge windows during its non-microlensing survey seasons, as well as observations from ground-based facilities like the Rubin Observatory, UKIRT, or PRime-focus Infrared Mirolensing Experiment (PRIME), could improve constraints on the microlensing parallax.}

\subsection{Impact of Birth Kicks on the $t_{E}$ Distribution}
\label{sec:BH tE birth kicks}
\begin{figure}
    \centering
    \includegraphics[scale=0.5]{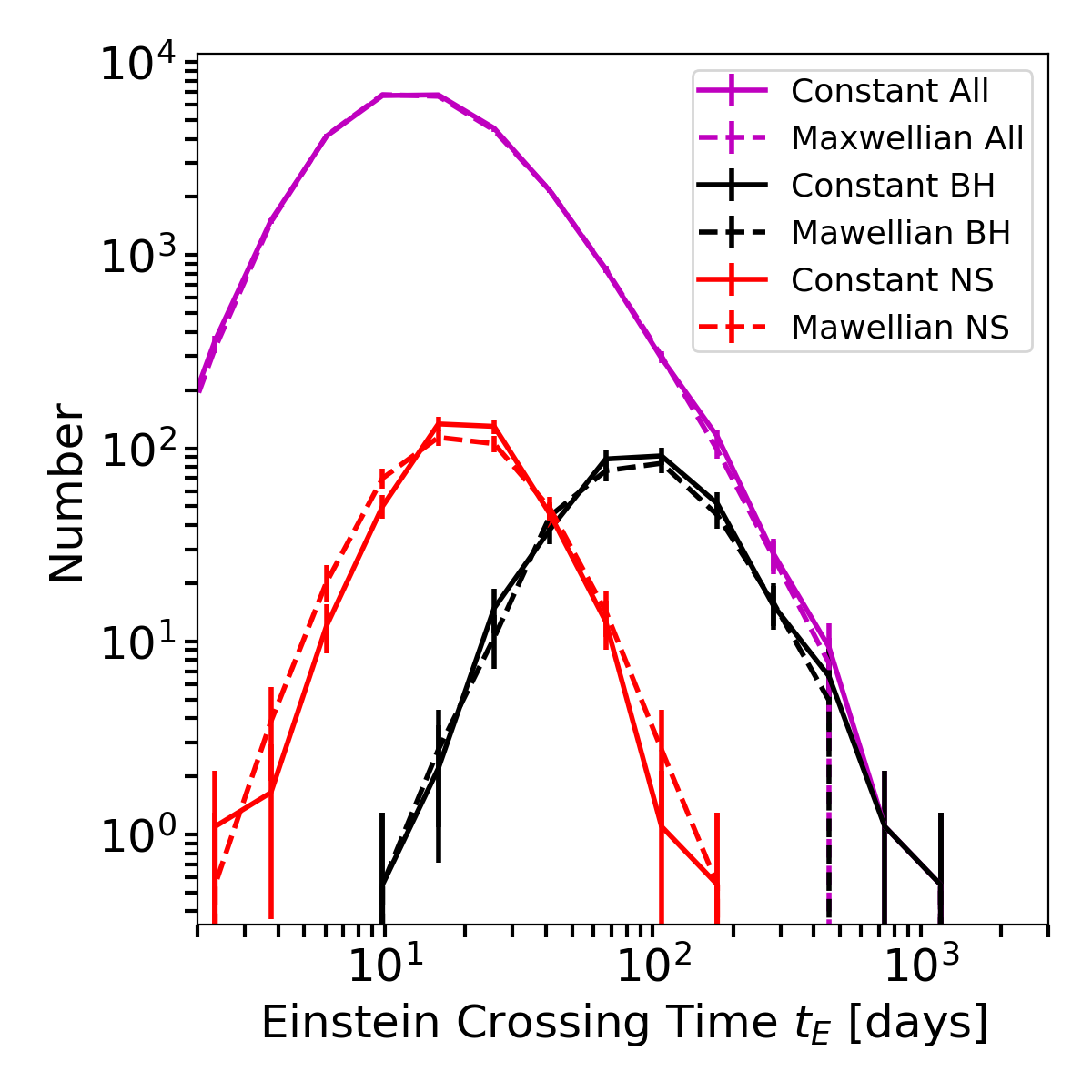}
    \caption{Einstein crossing time distributions for microlensing events detectable by a Roman style survey, for constant-valued versus Maxwellian NS/BH birth kicks.
    \edit{The number of events is scaled to the number of events Roman is expected to detect during its 5 year microlensing survey, as described in \S \ref{sec:Constraining the IFMR with the Roman Space Telescope}.
    The effect of the different birth kicks does not significantly change the $t_E$ distribution.}}
    \label{fig:birthkick_comp}
\end{figure} 

\edit{We next consider the effect of Maxwellian vs. single-valued birth kicks for NSs/BHs on the Einstein crossing time distribution.
For one of the Roman fields, OGLE-IV-BLG500, we run a PopSyCLE simulation identical to the Raithel18 IFMR simulation described in \S \ref{sec:Constraining the IFMR with the Roman Space Telescope}, except we use single-valued birth kicks.
The single-valued birth kick velocities are equal to the average velocity of the Maxwellian distribution, 350 km/s for NS and 100 km/s for BHs.
The comparison of the Einstein crossing time distribution between the two simulations are shown in Figure \ref{fig:birthkick_comp}.}

We find that the effect of Maxwellian as opposed to single-value birth kick velocities has a fairly weak impact on the distribution of Einstein crossing times.
\edit{This is because the kick velocities are added on top of existing stellar velocities to calculate the final remnant velocities (\S \ref{sec:NS/BH Birth Kick Velocities}).
Because there is already a significant amount of dispersion in the stellar velocity distribution, the effect of the additional dispersion from the Maxwellian kick distribution is diluted.
The $t_E$ distribution for NSs is slightly wider, with a longer tail toward short $t_E$, when using a Maxwellian distribution as compared to a single kick velocity.
However, within the uncertainties, the difference is not significant.
For BHs and the full lens population, the $t_E$ distributions are identical within the uncertainties.
This means that even with 20,000 photometric microlensing events, different kick velocity distributions with the same mean will likely be indistinguishable using Einstein crossing time distributions alone.}

\section{Discussion}
\label{sec:Discussion}

\subsection{Defining the Milky Way BH Microlensing Sample}
\label{sec:Defining the Milky Way BH Microlensing Sample}

From Figure \ref{fig:piE_tE_plot} it is apparent that, regardless of IFMR, BH lenses have the longest Einstein crossing times and the lowest microlensing parallaxes. 

\begin{figure}
    \centering
    \includegraphics[scale=0.5]{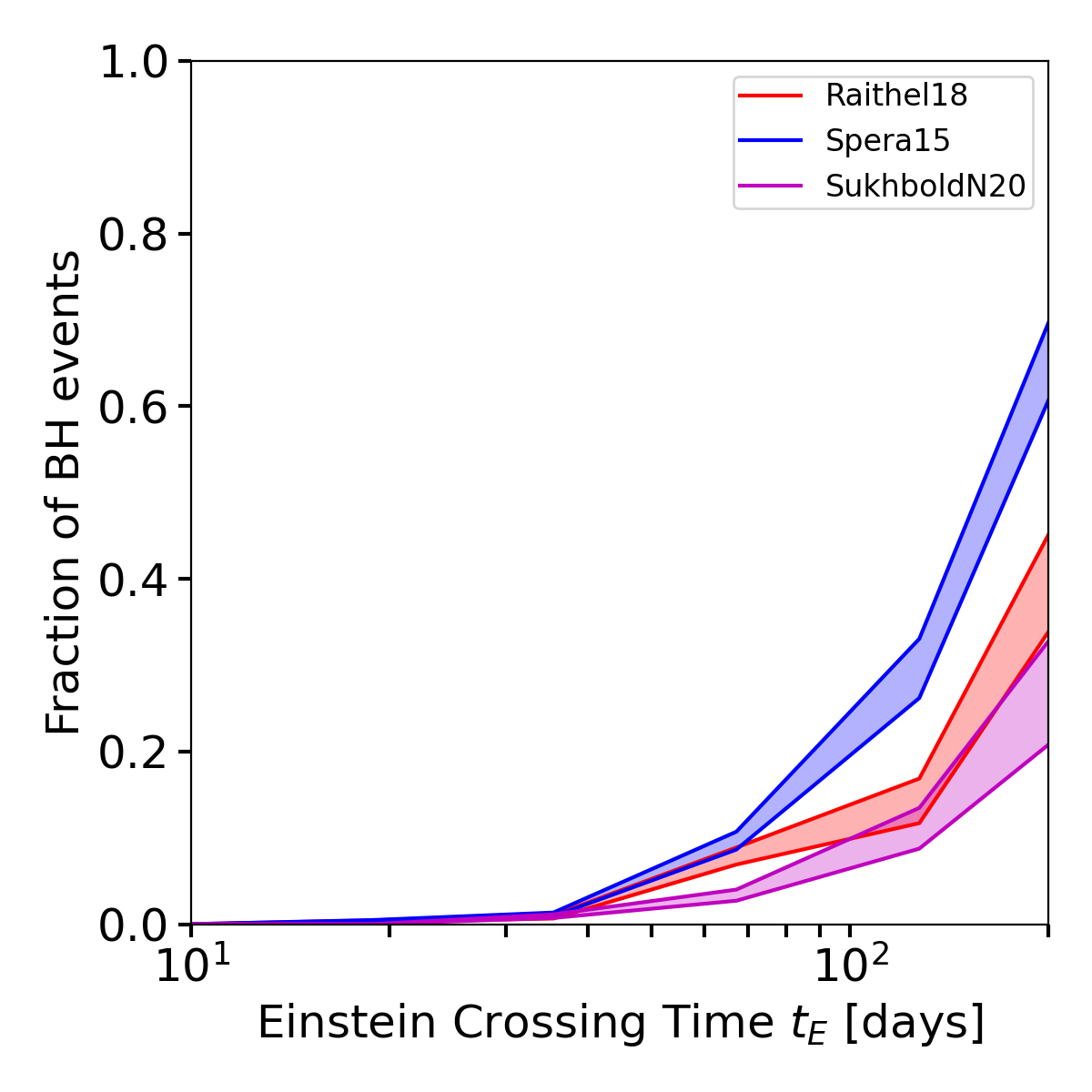}
    \caption{Fraction of BH events compared to total events in the OGLE-like survey sample as a function of $t_{E}$. 
    \edit{The number of events is scaled to the number of events an OGLE-like survey would observe over a span of 10 years, as described in \S \ref{sec:Constraining the IFMR with OGLE}.
    The SukhboldN20 IFMR produces the fewest BHs and has the lowest fraction of BH events at long $t_E$ ($>100$ days).
    The Spera15 IFMR produces the most BHs and has the highest fraction of BH events at long $t_E$. }
    Regardless of IFMR, roughly at least a third of events with $t_{E} >$ 150 days are BH lenses.}\label{fig:tE_frac}
\end{figure}

\begin{figure}
    \centering
    \includegraphics[scale=0.5]{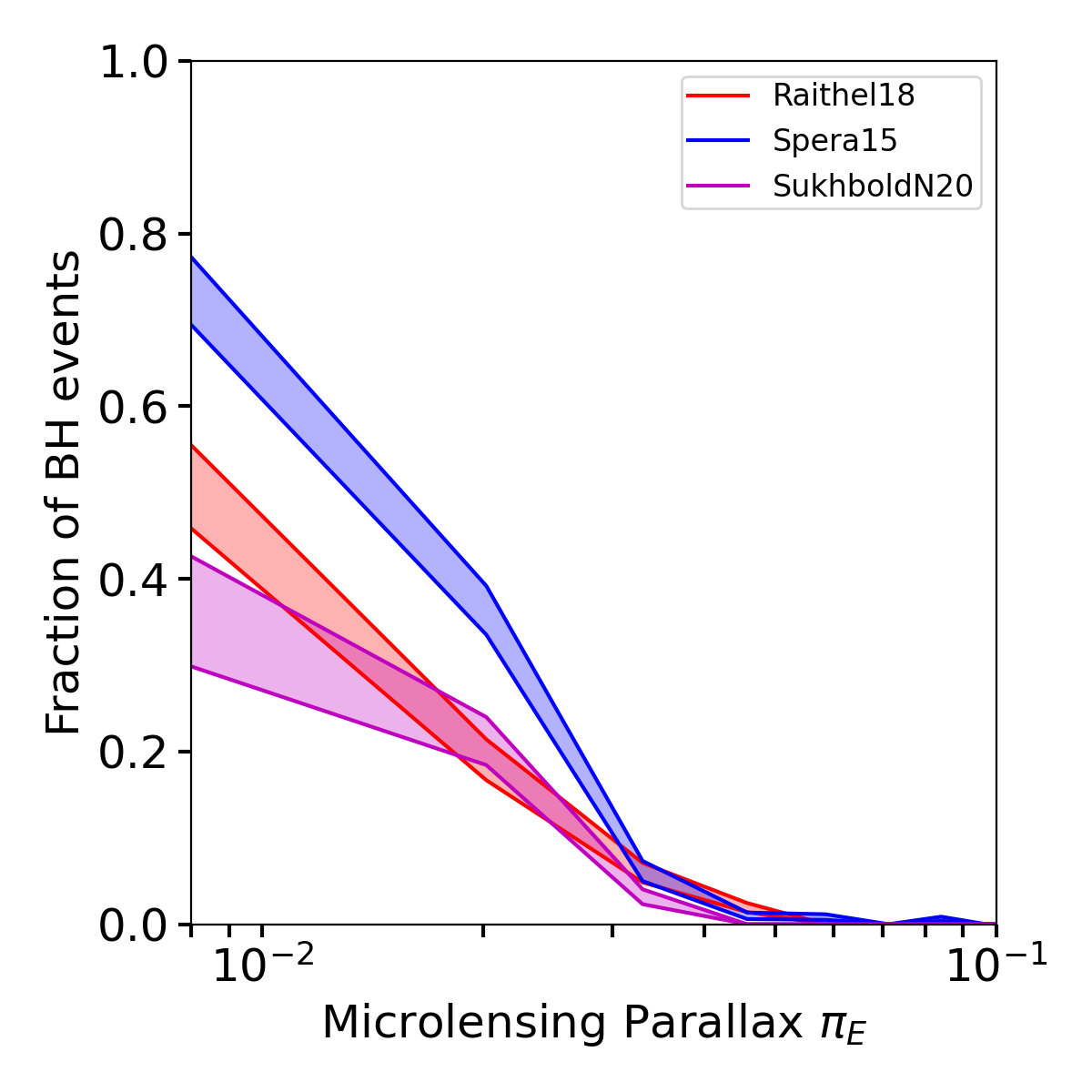}
    \caption{Fraction of BH events compared to total events in the OGLE-like survey sample as a function of $\pi_{E}$.
    \edit{The number of events is scaled to the number of events an OGLE-like survey would observe over a span of 10 years, as described in \S \ref{sec:Constraining the IFMR with OGLE}.
    Because the Spera15 IFMR produces more massive BHs than Raithel18 and SukhboldN20, it has the highest fraction of BH events at small $\pi_E$ ($<0.02$).
    Although the SukhboldN20 IFMR produces more massive BHs than the Raithel18 IFMR, it also produces fewer BHs overall, so the fraction of BH events at small $\pi_E$ for those two IFMRs are similar.}}\label{fig:piE_frac}
\end{figure}

\edit{Astrometric follow-up can be used to measure the lens masses of candidate BHs \citep{Lu:2016, Lam:2022, Sahu:2022}.}
As more astrometric follow-up campaigns are done it will be useful to quantify what percentage of events we should expect to be BH lenses based on our candidate selection criteria (either a minimum $t_{E}$ or a maximum $\pi_{E}$). 
Figures \ref{fig:tE_frac} and \ref{fig:piE_frac} show the ratio of BH events to total events for a 10 year OGLE-like survey, as a function of $t_{E}$ and $\pi_{E}$, respectively. 

Different IFMRs produce different BH fractions depending on the exact selection criteria.
For example, a dearth of BH lenses in a sample of candidates chosen for astrometric follow-up based on $t_{E} >$ 100 days might indicate that the Milky Way IFMR is closer to SukhboldN20, and that the BH to NS formation ratio is lower then predicted by Spera15. 

\subsection{Comparisons With Existing Compact Object Mass Distributions}
\label{sec:Comparisons With Existing Compact Object Mass Distributions}
Using metallicity distributions and a star formation history from \texttt{Galaxia}, we create a present-day black hole mass function (PDBHMF) for the Milky Way for each of the IFMRs. 
We find the total number of black holes produced by the Raithel18, Spera15, and SukhboldN20 IFMRs are $2.18 \times 10^{8}$, $2.67 \times 10^{8}$, and $1.72 \times 10^{8}$ respectively.
Current estimates of the total number of stellar mass black holes in the Milky Way range between $10^{8}-10^{9}$ \citep{Agol:2002}.

\begin{figure}
    \centering
    \includegraphics[scale=0.5]{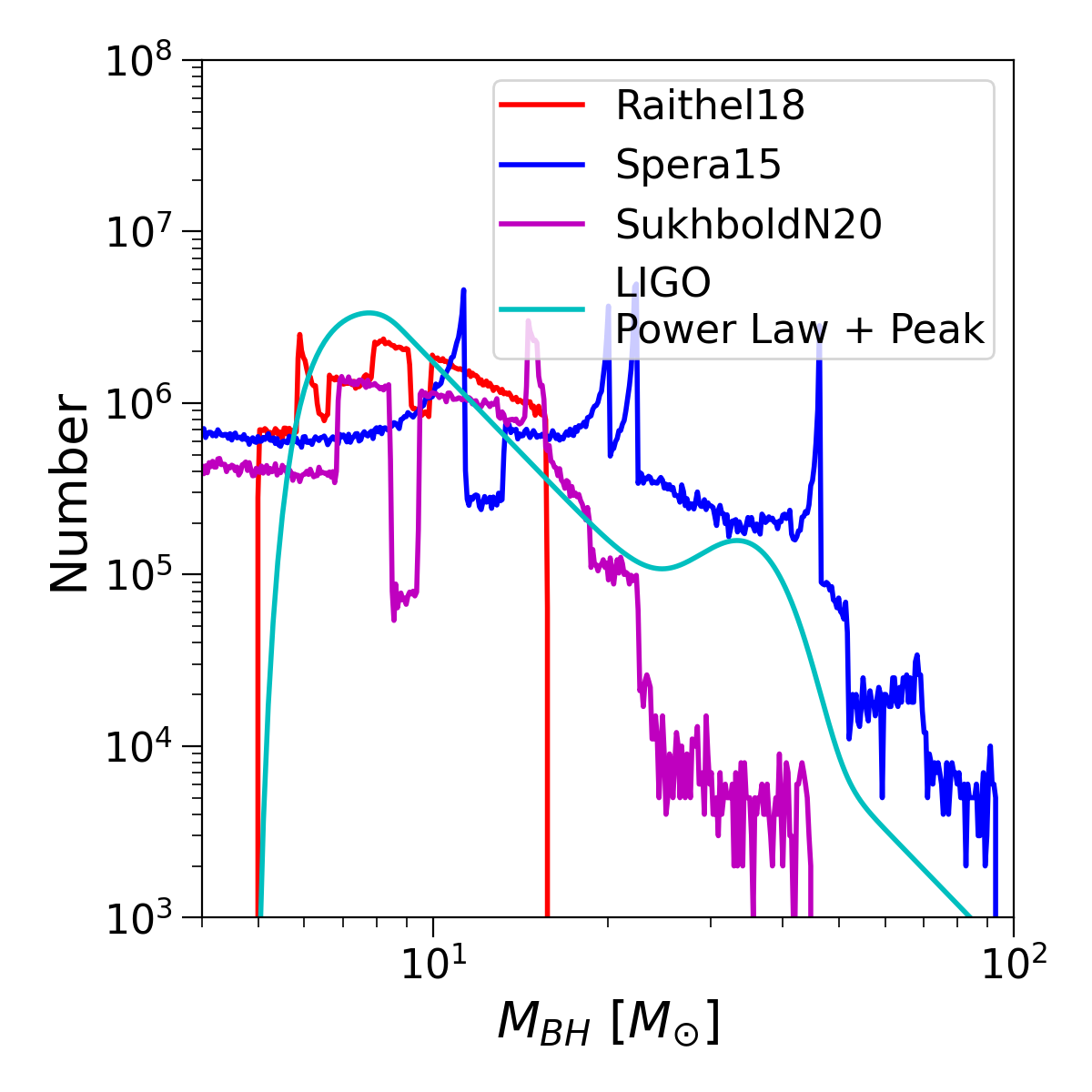}
    \caption{Comparisons of the present day black hole mass function of the Milky Way produced using star formation history and metallicity from the \texttt{Galaxia} simulation and the Raithel18, Spera15 and SukhboldN20 IFMRs of SPISEA with the astrophysical black hole mass function generated from \edit{LIGO GWTC-3 results \citep{LIGO:2021_gwtc3} using their Power Law + Peak model \citep{LIGO:2021_pop_props}.}
    \edit{The four spikes in the Spera15 IFMR present day black hole mass function are artifacts of the metallicity binning scheme in the population synthesis procedure described in \S \ref{sec:Metallicity Binning}.
    Finer metallicity bins would remove these artificial spikes at the expense of increased computational time.}}
    \label{fig:LIGO_comp}
\end{figure}

\edit{Figure \ref{fig:LIGO_comp} compares the Milky Way PDBHMF predicted by each IFMR against the fiducial ``Power Law + Peak" binary black hole (BBH) primary mass distribution derived using Gravitational Wave Transient Catalog 3 \citep[GWTC-3,][]{LIGO:2021_gwtc3}.
For comparison purposes, the ``Power Law + Peak" model has been vertically scaled so that its maximum roughly matches the maxima of Galactic BH distributions produced by the IFMRs.
The GWTC-3 BBH population is not a direct analogue to the simulated isolated Milky Way BH population, as the former is an extragalactic binary population at low metallicity.
However, a comparison of the two populations' similarities and differences is still worthwhile.}

Note that the four large spikes in the Spera15 IFMR PDBHMF are artifacts of the coarse metallicity bins used in this simulation (see \S \ref{sec:Metallicity Binning} for more details).
Finer metallicity bins would smooth over these spikes at the expense of increased computational time and the need to store many more \texttt{SPISEA} isochrones. 

\edit{None of the PDBHMFs predicted by the different IFMRs match the fiducial BBH Power Law + Peak distribution.
All three IFMRs have approximately flat PDBHMFs for masses $\lesssim 20 M_\odot$, while the BBH primary mass distribution follows a power law with spectral index $\alpha \sim -3.5$ \citep{LIGO:2021_gwtc3}.
None of the IFMRs reproduce the peak at $35 M_\odot$ in the BBH mass distribution.}

\edit{The slope of the Spera15 IFMR PDBHMF is in reasonable agreement with the Power Law + Peak model for masses $\gtrsim 35 M_{\odot}$.
This is likely because the high mass end of the Spera15 IFMR PDBHMF is composed of high mass BHs that came from a low metallicity progenitor population, similar to the progenitors of the GWTC-3 BBH population.
In contrast, the low mass end of the PDBHMF does not match.
The SukhboldN20 IFMR PDBHMF has a similar slope to the power law portion of the Power Law + Peak model for masses $\gtrsim 15 M_\odot$, modulo the peak at $\sim 35 M_\odot$.
Similar to the explanation for the Spera15 IFMR, the progenitors of the BHs in this mass range have low metallicity progenitors
The Raithel18 IFMR PDBHMF does not have the massive BHs in the BBH population as all progenitors are assumed to be solar metallicity.}

Both the Raithel18 IFMR and the Power Law + Peak model favor a minimum BH mass above 5 $M_{\odot}$, while both Spera15 and SukhboldN20 IFMRs allow for BH masses down to 3 $M_{\odot}$.
There is observational evidence for mass gap BHs between $2 - 5 M_\odot$ including two LIGO merger remnants, one at $\sim 3 M_\odot$ \citep{GW170817} and another at $\sim 3.4 M_\odot$ \citep{GW190425}, and the $\sim 2.6 M_\odot$ merger component of GW190814 \citep{GW190425}.
There has also been a detection of a $\sim 3 M_\odot$ BH \edit{in a non-interacting binary system} \citep{Thompson:2019}. 
These results indicate that the IFMR should allow for at least a few BHs below $\sim 5 M_\odot$.

\subsection{Future Work}
\label{sec:Future Work}

\subsubsection{Mass Dependent Birth Kicks}
\label{sec:Mass Dependent Birth Kicks}
\edit{In this work we updated the way birth kicks are applied to NSs and BHs in \texttt{PopSyCLE}.
Previously, each population was assigned a single kick velocity value.
Now, the kick values for each population are drawn from a Maxwellian distribution.}
A more accurate implementation would also have the birth kick velocities \edit{be a function of} mass.
In future versions of \texttt{PopSyCLE} we \edit{will consider different kick formalisms, such as having kick velocity decreasing with increasing remnant mass, or implementing} predictions for birth kick velocities coming from gravitational wave observations \citep{O'Shaughnessy:2017}. 
\edit{If more massive BHs were given smaller birth kicks, this would likely make the Einstein crossing times slightly longer on average for massive BHs, and slightly shorter for less massive BHs.}

\subsubsection{Predicting $\pi_{E}$}
\label{sec:fisher matrix stuff}
The best BH candidates obtained from photometric microlensing surveys have both long Einstein crossing times and low microlensing parallaxes. 
Up until this point BH candidates for astrometric follow-up have mainly been selected based on $t_{E}$ alone \citep{Lu:2016, Lam:2022, Sahu:2022}. 
However, even for $t_{E} > 120$ days, around half of lenses are not BHs (Figure \ref{fig:tE_frac}).
Selecting on microlensing parallax in addition to Einstein crossing time would improve our ability to pick BH candidates for astrometric follow-up. 
However, it is unclear how well $\pi_{E}$ can be predicted prior to the photometric peak. 

It \edit{may be} possible to use Fisher matrix analysis to constrain $\pi_{E}$ over the course of the microlensing event, \edit{although it will be difficult in the case of BH lenses which have small $\pi_E$} \citep{Karolinski:2020}. 
One of the next steps motivated by this work will be to determine exactly how well $\pi_{E}$ can be predicted prior to peak photometric amplification, in order to determine the viability of $\pi_{E}$ for use as a BH candidate selection criteria.

\subsubsection{Binaries}
\label{sec:Binaries}
In this work we have assumed that all stars and compact objects are single.
\edit{However, a significant fraction of stars are known to be in binary systems, which can also act as sources or lenses.}
\edit{Large microlensing catalogs such as \citet{Mroz:2019} select only for microlensing events with a single lens and source.}
In some cases it is very clear that a microlensing signal is coming from a binary lens \edit{or source due to caustic crossings, significant asymmetry, or multiple peaks in the lightcurve.
However, some lensing geometries may result in lightcurves that resemble single lens and source events.
In other cases, observational gaps or data quality are not sufficient to rule out the possibility of binary microlensing.} 
Abrams et al. in preparation explores the effect of adding binaries to the \texttt{PopSyCLE} microlensing simulation. 

In addition to changes in the microlensing source and lens population, binarity can affect the IFMR itself.
All the IFMRs considered in this paper come from models of single star evolution, and so are only valid for single stars or binaries sufficiently wide enough to be non-interacting. 
Close binaries can allow for mass transfer between a massive star and its companion during the late stages of stellar evolution \citep{Ivanova:2013}. 
\edit{Some binary star IFMRs simply assume hydrogen envelope loss for close binaries \citep{Woosley:2020}, while others actually calculate the amount of mass and angular momentum transfer \citep{Yoon:2010}.
Population synthesis codes such as \texttt{StarTrack} \citep{Belczynski:2002, Belczynski:2008} that include binary stellar evolution have been used to investigate binary black hole mergers \citep{Belczynski:2016}. 
However, these have not been incorporated into microlensing simulations.
A future version of \texttt{PopSyCLE} will incorporate the results of binary stellar evolution.}

\section{Conclusions}
\label{sec:Conclusions}
In this work we explored the possibility of using photometric microlensing to constrain the IFMR for massive 
stars.
\begin{itemize}
    \item We have added two metallicity dependent IFMRs, Spera15 and SukhboldN20, to the \texttt{SPISEA} \edit{simple stellar} population synthesis code. 
    
    \item Different IFMRs yield different Galactic BH mass distributions.
    This in turn affects the observed distributions of microlensing event parameters.
    \edit{We focus on long $t_E$ and small $\pi_E$ in this work as these are the regimes most sensitive to BH lenses.}
    
    \item \edit{Considering the number of events with $t_E > 100$ \textcolor{black}{days} observed with an OGLE-like survey, the difference between the Spera15 and SukhboldN20 IFMRs is not statistically significant. 
    Considering the number of events with $\pi_E < 0.02$ observed with an OGLE-like survey, the difference between the IFMRs is statistically significant.
    However, ground-based surveys like OGLE are not able to constrain such small $\pi_E$, making it not possible to distinguish the IFMRs photometrically.}
    
    \item \edit{Considering the number of events with $t_E > 100$  \textcolor{black}{days} observed with a Roman-like survey, the difference between the Spera15 and SukhboldN20 IFMRs is statistically significant. 
    Considering the number of events with $\pi_E < 0.02$ observed with an OGLE-like survey, the difference between IFMRs is also statistically significant.
    Roman's excellent photometric precision should allow it to constrain small $\pi_E$, but this is tempered by the large observational gaps in the survey.
    Further work to determine how best these gaps can be filled, either by Roman itself or by other facilities, will maximize the ability of Roman to detect small $\pi_E$, and in turn constrain the IFMR and detect BHs.}
    
    \item \edit{Microlensing parallax is more sensitive to changes in lens mass than the Einstein crossing time.}
    Because black holes are high mass, this means that it will be important that future surveys be able to accurately measure small microlensing parallax signals ($\pi_{E} < 0.02$) in order to place meaningful constraints on the BH lens mass distribution. 
    
    \item Comparing astrometric BH lens detection efficiency based on candidate selection criteria to our predictions for each IFMR will be another way to place constraints on the Milky Way IFMR.
\end{itemize}

\section{Acknowledgements}
We thank Tuguldur Sukhbold for providing the models used in the SukhboldN20 IFMR and helping with the interpolation scheme.
\edit{We thank the referee for feedback that improved the results of this paper.}
J.R.L. and C.Y.L. acknowledge support by the National Science Foundation under Grant No. 1909641 and the National Aeronautics and Space Administration (NASA) under contract No. NNG16PJ26C issued through the WFIRST (now Roman) Science Investigation Teams Program.
C.Y.L. also acknowledges support from NASA FINESST grant No. 80NSSC21K2043.
M.W.H. acknowledges support by the Brinson Prize Fellowship

\emph{Software}: \texttt{Galaxia} \citep{Sharma:2011}, \texttt{SPISEA} v2.1 \citep{Hosek:2020}, \texttt{PopSyCLE} \citep{Lam:2020}, MIST v1.2 (\citet{Choi:2016} and \citet{Dotter:2016}), Astropy (\citet{Astropy:2013} and \citet{Astropy:2018}), Matplotlib \citep{Hunter:2007}, NumPy \citep{Walt:2011}, SciPy \citep{Virtanen:2019}. 

\bibliographystyle{aasjournal}
\bibliography{main.bib}

\appendix

\section{Neutron Star Mass Distribution}
\label{sec:Neutron Star Mass Distribution}
The Raithel18 IFMR (described in \S \ref{sec:The Raithel18 IFMR Function}) and SukhboldN20 IFMR (described in \S \ref{sec:Sukhbold N20 IFMR}) as implemented in this work do not use the NS masses predicted by the papers from which they were drawn.
What is instead used is the mass dependent probability (mass and metallicity dependent for SukhboldN20) that a given progenitor will form a BH or a NS. 
Once an object is determined to be a NS, it is assigned a mass drawn from a Gaussian distribution with $\mu$ = 1.36 $M_{\odot}$ and $\sigma$ = 0.09 $M_{\odot}$. 
This distribution is based on the observed masses of pulsars in binaries.

\begin{figure*}
    \begin{center}
    \includegraphics[width=\textwidth]{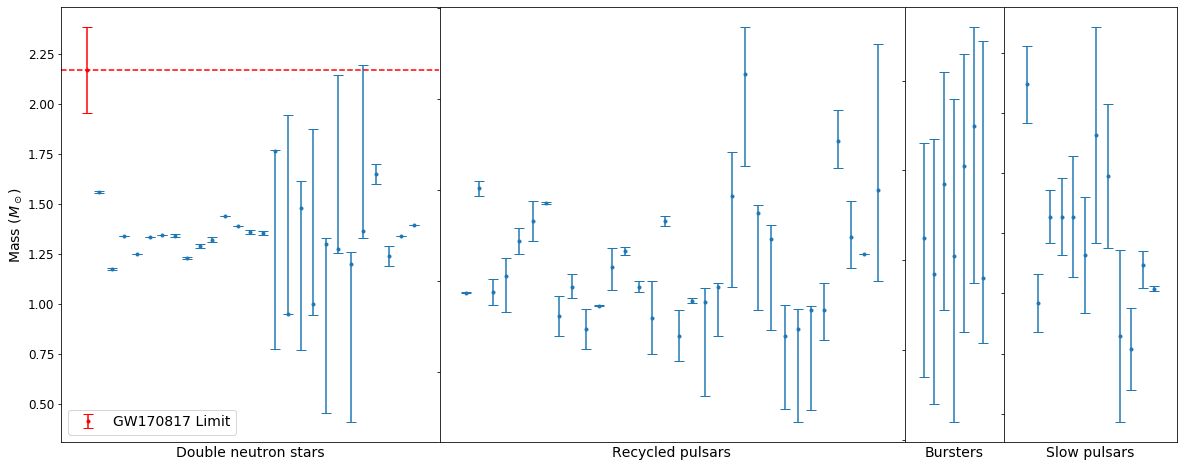}
    \caption{Inferred neutron star masses for source type using relativistic pulsar timing and Doppler spectroscopy. The upper mass limit obtained from the DNS GW170817 is shown in red \citep{Margalit:2017}\label{fig:NS_masses}. Error bars indicate a 90\% confidence interval.}
    \end{center}
\end{figure*}

Although there have been on the order of $10^3$ pulsars observed, only $10\%$ are in binary systems \citep{Abdo:2013}. 
Therefore, the current techniques relying on extracting mass information from orbital motion of the neutron star obtain a fairly small subset of the entire population. 
Despite this, precise Neutron Star (NS) mass measurements have been made using a variety of sources: Double Neutron Stars (DNS), Recycled Pulsars (RP), Bursters, and Slow Pulsars (SP). 
We add some newly discovered pulsars to the sample already included in \citet{Ozel:2016} --- J1811+2405 \cite{Ng:2020}, J2302+4442 \cite{Kirichenko:2018}, J2215+5135 \cite{Linares:2018}, J1913+1102 \cite{Ferdman:2018}, J1411+2551 \cite{Martinez:2017}, J1757+1854 \cite{Cameron:2018}, J0030+0451 \cite{Riley:2019}, J1301+0833 \cite{Romani:2016}. 
We plot this sample in Figure \ref{fig:NS_masses}, including the upper mass limit inferred from the GW170817 merger for reference only \citep{Margalit:2017}. 
Using this sample we obtain a fit for the NS mass distribution using a Bayesian MCMC method adapted from \citep{Kiziltan:2010}.
Assuming a Gaussian distribution of pulsars, the authors use the Metropolis-Hastings MCMC \citep{Hastings:1970} algorithm to obtain the mean mass for their samples of DNS and NS-WD system measurements. 
A similar algorithm is used in \cite{Ozel:2016} and other papers on the NS mass distribution.

We assume that the that the NS mass distribution is Gaussian and quantify the mass errors, approximating them also as Gaussian distributions, and are able to specify our prior beliefs about $\mu$ and $\sigma$ (equations \ref{eq:avg_prior} and \ref{eq:sig_prior} respectively). 
For the measurements with non-symmetric errors, we take the larger of the two error values to be $1\sigma$.
For a more thorough treatment of this problem, we would fit a more accurate probability distribution to the error values. but that is beyond the scope of this paper.

Here, we summarize the statistical method adopted from \cite{Kiziltan:2010}. 
This method attempts to model the posterior of the measured NS data through a randomized Markov-Chain-Monte-Carlo (MCMC) method.
It is particularly applicable in problems in which there is a small amount of data available, as is the case here (26 DNSs, 32 RPs, 7 Bursters, 12 SPs). 
When applied correctly, MCMC is able to accurately the uncertainty of its own results even when the sample size is small. 
We model the neutron star distribution as a Gaussian with some mean and standard deviation. The likelihood of this distribution is then described by:

\begin{align}
\mathcal{L}(\mu,\sigma^2|\text{data}) &= \frac{P(\text{data}|\mu,\sigma^2) P(\mu)P(\sigma^2)}{P(\text{data})}\\
&\propto \mathcal{L}(\text{data}|\mu,\sigma^2) \pi(\mu)\pi(\sigma^2)\\
\mathcal{L}(\text{data}|\mu,\sigma^2) &= \prod_i^{N}[2\pi(\sigma^2+S_i^2)]^{-1/2}e^{-\frac{(m_i-\mu)}{2(\sigma^2+S_i^2)}} \\
\end{align}

Where $m_i$ and $S_i$ are the values for the mass and mass error of measurement $i$, respectively. 
The prior distributions for the parameters $\mu$ and $\sigma$, also adopted from \cite{Kiziltan:2010}, are modeled as Normal (\ref{eq:avg_prior}) and Inverse Gamma (\ref{eq:sig_prior}) distributions, respectively. 
These priors are tuned with the hyperparameters a, b, c, and d, as shown below:

\begin{align}
\pi(\mu) &= N(a, b^2) = (2\pi b^2)^{-1/2}\text{exp}\left[-\frac{(\mu-a)^2}{2b^2} \right] \label{eq:avg_prior}\\
\pi(\sigma^2) &= \Gamma^{-1}(c, d) = \frac{d^c e^{-d/\sigma^2}}{\Gamma(c)\sigma^{2(c+1)}}\label{eq:sig_prior}
\end{align}

Since a more rigorous search for the best hyperparameters is beyond the scope of this paper, we use the best results from \cite{Kiziltan:2010}, which are as follows: $a=1.4$, $b=0.05$, $c=5$, $d=0.01$. 
As much of our data comes from this paper, the best values for our data should not vary significantly from these. The full expression for the posterior distribution (in log form) is shown below:

\begin{align}
    \log(\mathcal{L}(\mu,\sigma^2|\text{data})) &= -\frac{1}{2} \sum_{i}^N \frac{(m_i-\mu)^2}{\sigma^2+S_i^2} + \log(\sigma^2+S_i^2) + \log(\pi(\mu)) + \log(\pi(\sigma)) \label{eq:post}
\end{align}

The algorithm generates values of $\mu$ and $\sigma$ from their prior distributions, and compares the posteriors of the results using equation \ref{eq:post}. 
If a $\mu,\sigma$ combination is more likely than the current one, the step will be accepted, and if not it will be rejected with some probability related to the likelihood. 
If the data is modeled correctly, the algorithm will converge to the true posterior distribution.

\section{Galactic Model Comparisons}
\label{sec:Galactic Model Comparisons}

\begin{deluxetable*}{lrr|rrr|rrr}
\tablecaption{\texttt{PopSyCLE} vs.~Observed Event Rates for Different Galactic Models
    \label{tab:PopSyCLE Galactic Model Comparison}}
\tablehead{
  \colhead{Name} & 
  \colhead{$l$} &
  \colhead{$b$} &
  \multicolumn{3}{c}{$n_{*}$ ($10^{6}$)} &
  \multicolumn{3}{c}{$\Gamma$ ($10^{-6}$)} \\
  \colhead{} &
  \colhead{(deg)} &
  \colhead {(deg)} &
  \multicolumn{3}{c}{(stars deg$^{-2}$)} &
  \multicolumn{3}{c}{(events star$^{-1}$ yr$^{-1}$)} \\
  \colhead{} &
  \colhead{} &
  \colhead{} &
  \colhead {\citetalias{Mroz:2019}} &
  \colhead{v2} &
  \colhead{v3} &
  \colhead{\citetalias{Mroz:2019}} &
  \colhead{v2} &
  \colhead{v3} \\
  \colhead{} &
  \colhead{} &
  \colhead{} &
  \colhead {} &
  \colhead{(Sim.)} &
  \colhead{(Sim.)} &
  \colhead{} &
  \colhead{(Sim.)} &
  \colhead{(Sim.)}}
\startdata
OGLE-IV-BLG500 & 1.00 & -1.03 & 4.84 & 4.75 & 3.37 & 23.9 $\pm$ 2.0 & 60.1 $\pm$ 3.7 & 31.6 $\pm$ 3.2 \\ 
OGLE-IV-BLG506 & 0.01 & -3.00 & 9.19 & 5.93 & 3.83 & 16.5 $\pm$ 1.1 & 53.1 $\pm$ 3.1 & 18.0 $\pm$ 2.2 \\ 
OGLE-IV-BLG675 & 0.78 & 1.69 & 4.03 & 5.76 & 3.94 & 26.5 $\pm$ 2.3 & 63.7 $\pm$ 3.4 & 22.6 $\pm$ 2.5 \\ 

\enddata
\tablecomments{\edit{Observed vs. simulated stellar density and efficiency-corrected event rates for three select fields in the OGLE-IV survey representative of the Roman microlensing fields.
The observed stellar density and event rates are calculated as described in Table \ref{tab:PopSyCLE Fields}. 
Here, we consider two different Galactic bar models.
In the v2 Galactic model, $\alpha$ = 11.1$^{\circ}$ and $x_{0}$ = 1.59 kpc.
In the v3 Galactic model, $\alpha$ = 28$^{\circ}$ and $x_{0}$ = 0.7 kpc.
Both the v2 and v3 models here use the SukhboldN20 IFMR.
We only consider a single IFMR as the difference in event rate across different IFMRs (Table \ref{tab:PopSyCLE Fields}) is smaller than the difference of changing the Galactic model.}}
\end{deluxetable*}

\edit{We investigate how Galactic model uncertainties propagate to the predictions made  by microlensing simulations by focusing on the effect of one particularly uncertain aspect: the geometry of the Galactic Bar.
We consider two different Galactic models denoted ``v2" and ``v3", described in detail in Appendix A of \citet{Lam:2020}.
In short, the models differ in the angle from the line connecting the Sun and Galactic Center $\alpha$ and major axis length of the Galactic bar $x_0$. 
In the v2 Galactic model, the bar is longer and less tilted along our line of sight, with $\alpha$ = 11.1$^{\circ}$ and $x_{0}$ = 1.59 kpc.
In the v3 Galactic model, the bar is shorter and more tilted along our line of sight, with $\alpha$ = 28$^{\circ}$ and $x_{0}$ = 0.7 kpc.}

\edit{In the main body of this paper, all simulations were run using the v3 Galactic model as it better matched the event rate/star/year presented in \citet{Mroz:2019} than v2.
In this Appendix, we investigate how different the outcomes of our simulations are if we instead use the v2 model.}

\edit{In Table \ref{tab:PopSyCLE Galactic Model Comparison} we compare simulated completeness-corrected event rates and stellar densities for three fields in the OGLE IV survey to results from \citetalias{Mroz:2019}, following the methodology described in the first paragraph of \S \ref{sec:Comparison against observations}.
The simulated values all use the SukhboldN20 IFMR; we only consider a single IFMR in this Appendix as the difference in event rate across different IFMRs (Table \ref{tab:PopSyCLE Fields}) is smaller than the difference of changing the Galactic model.
Consistent with the results in Appendix A of \citet{Lam:2020}, the v2 Galactic model results in an event rate significantly higher than that of \citet{Mroz:2019} but with improved agreement in the stellar density.}

\edit{Next, we consider the number of observable microlensing events in an OGLE-like or Roman-like microlensing survey. 
We calculate the number of events in the three fields using the methodology described in \S \ref{sec:BH Microlensing Statistics} and \S \ref{sec:Constraining the IFMR with the Roman Space Telescope} for OGLE and Roman, respectively.
In both cases, there are are about 3.5$\times$ more observed events when using the v2 Galactic model than when using the v3 Galactic model which makes the IFMR more easily distinguishable.}
 
\edit{Ultimately, for the conclusions presented in the main body of the paper, we are solely interested in whether photometric microlensing observations can distinguish between different IFMRs, \emph{given} a Galactic model; we assume that the Galactic model is constrained by other means (e.g. star counts, proper motion measurements). 
Whether or not photometric microlensing observations alone can simultaneously constrain the IFMR and Galactic model are beyond the scope of this work.}

\end{document}